\documentclass[aps,superscriptaddress,eqsecnum,nofootinbib,preprintnumbers]{revtex4}
\pdfoutput=1
\usepackage{graphicx,epsfig}
\usepackage{amsmath}
\usepackage {amssymb}
\usepackage{placeins}

\begin{document}

\title{Generalized Non-Minimal Derivative Coupling: Application to Inflation and Primordial Black Hole Production}

\author{Ioannis Dalianis}
\email{dalianis@mail.ntua.gr} \affiliation{Physics Division,
National Technical University of Athens, 15780 Zografou Campus,
Athens, Greece.}

\author{Stelios Karydas }
\email{stkarydas@mail.ntua.gr} \affiliation{Physics Division,
National Technical University of Athens, 15780 Zografou Campus,
Athens, Greece.}

\author{Eleftherios Papantonopoulos}
\email{lpapa@central.ntua.gr} \affiliation{Physics Division,
National Technical University of Athens, 15780 Zografou Campus,
Athens, Greece.}


\begin{abstract}
\vspace{5mm}
We introduce inflationary models where the inflaton features a field dependent non-minimal derivative coupling to the Einstein tensor, that we name GNMDC. This  Horndeski term gives new and distinguishable inflationary predictions in a  framework that ameliorates
possible problems related with gradient instabilities during the reheating stage.  We examine the inflationary phenomenology using power law monomial and exponential potentials. We further elaborate on the implications of the inflaton-modulated GNMDC and construct GNMDC terms that can  amplify the power spectrum of primordial perturbations at small scales, triggering PBH production. An attractive feature of  the GNMDC is that inflation, as well as PBH production, can be implemented utilizing the Higgs potential.
\end{abstract}
\maketitle

\section{Introduction}

The inflationary description of the early universe
offers a compelling explanation for the initial conditions of the hot big bang.
In the inflationary scenario an early accelerating expansion takes place that can be viewed as the gravitational effect of the inflaton field itself.
Nowadays, precision cosmology tests the inflationary paradigm
and, at the same time, the gravity that operates at very high densities. The identification of a viable inflationary model requires
the full investigation of the inflaton-gravity dynamics that can be achieved only after  a survey of the possible modifications of the gravity theory has been carried out.

The standard theory of General Relativity (GR) is modified when extra higher-order geometric terms of high curvature are introduced in the action, or extra scalar fields are present which are non-minimally coupled to gravity. Higher-order corrections to the Einstein-Hilbert term  arise naturally in the gravitational effective action of String Theory \cite{Gross:1986mw}.
On the other hand, the introduction of extra scalar fields results to scalar-tensor theory \cite{Fujii:2003pa}.
A particularly studied scalar-tensor theory is the one resulting from the Horndeski Lagrangian
\cite{Horndeski:1974wa}. Horndeski theories are manageable since they yield second-order field equations and do not produce ghost instabilities \cite{Ostrogradsky:1850fid}. Moreover, many scalar-tensor theories that modify GR
 share a classical Galilean symmetry \cite{Nicolis:2008in,Deffayet:2009wt,Deffayet:2009mn,Deffayet:2011gsz,Kobayashi:2011nu,Kamada:2010qe}.
Horndeski theory, includes, among other terms, the
non-minimal derivative coupling of the scalar field to the Einstein tensor (NMDC).
The NMDC coupling  has interesting implications both on short and large distances for black hole physics  \cite{Kolyvaris:2011fk, Rinaldi:2012vy, Kolyvaris:2013zfa, Koutsoumbas:2015ekk} and inflation \cite{Amendola:1993uh} respectively. For a recent review, see \cite{Papantonopoulos:2019eff}.

 From the inflationary model building point of view, the attractive feature is that the non-minimal derivative coupling acts as a friction mechanism allowing steep potentials to implement a slow-roll phase  \cite{Amendola:1993uh,Sushkov:2009hk} and inflation with potentials such as Standard Model Higgs can be realized \cite{Germani:2010gm}. Additionally,  inflationary potentials within the NMDC framework can be consistently described in supergravity  \cite{Farakos:2012je, Farakos:2013zya}  via the gauge kinematic function \cite{Dalianis:2014sqa}. 
 The inflationary predictions  of the NMDC  were fully investigated in \cite{Dalianis:2016wpu} where 
the dynamics of both the inflationary slow-roll phase and the subsequent reheating were considered.

Actually, the presence of the NMDC modifies the standard picture of the reheating phase.
Due to the NMDC the inflaton oscillates rapidly without significant damping \cite{Sadjadi:2012zp, Ghalee:2013ada, Gumjudpai:2015vio, Myung:2016twf, Ema:2015oaa, Ema:2016hlw} and affects  heavy particle production \cite{Koutsoumbas:2013boa}.
The implication of such oscillations, where the NMDC dominates over the canonical kinetic term, might be problematic for
the stability of the postinflationary system, due to the oscillations of the sound speed squared between positive and negative values \cite{Ema:2015oaa}.
This implies that the scalar fluctuations are exponentially enhanced, particularly the shortest wavelength mode of the scalar perturbations.
The gravity-inflaton system soon becomes non-linear, and the dynamics are  difficult to follow  analytically  invalidating the inflationary predictions of the model, with particular exceptions such as the new Higgs inflation  \cite{Germani:2015plv}.

The instability can be avoided when the non-minimal kinetic term is negligible, when compared to the canonical one, during the oscillatory period.
However, when this condition is satisfied, the model effectively reduces to just a canonical scalar field with Einstein gravity even during the inflationary period, and the advantages of the NMDC are lost. These facts discourage the use of the simple NMDC version for inflationary model building. However, the NMDC studied so far is a special case of the Horndeski Lagrangian density \cite{Deffayet:2011gsz,Kobayashi:2011nu}
\begin{align} \label{lagran}
{\cal L}_5=G_5(\phi, X) G^{\mu\nu}\partial_\mu \phi \partial_\nu \phi \,~
\end{align}
where $X=\partial_\mu \phi \partial^\mu \phi /2$.
The function $G_5(\phi, X)$ is usually chosen to be a constant function, $G_5(\phi, X)=1/M^2$. Though this choice is the simplest, it is the reason behind the problematic postinflationary evolution.
Instead, if one chooses a more general function $G_5(\phi, X)=f(\phi)\, \xi(X)$, that we call {\it general non-minimal derivative coupling} (GNMDC) the phenomenology of the Horndeski terms becomes richer, both during inflation and reheating stages.
For $f(\phi)\propto \phi$ the GNMDC term vanishes each time the inflaton field crosses the minimum of the potential  and  we find that the system, after a few oscillations, transits to the dynamics of a canonical scalar field with Einstein gravity.
Hence, thanks to the GNMDC, inflationary models turn to being calculable and reliable, dominated by GR dynamics during the reheating stage.
In this work we examine the inflationary phenomenology focusing on  the {\it  Higgs potential}, which is known to exist in nature  \cite{Chatrchyan:2012xdj} and on {\it exponential potentials}, which are motivated by several beyond the Standard Model theories.

An extra motivation for introducing the GNMDC is the recent results from the  LIGO Scientific and Virgo Collaborations \cite{Abbott:2016blz,Abbott:2016nmj}, which provided a very tight bound on the speed of gravitational waves (GWs). 
Assuming that a scalar field with NMDC plays the role of dark energy it was  found \cite{Germani:2010gm,Germani:2011ua} that
the propagation speed of the tensor perturbations in a Friedmann-Robertson-Walker cosmological background
is different from the speed of light $c$. The measurement of the speed of the GWs, constrains deviations (up to order of $10^{-15}$)  from the speed of light,
and it was argued that dark energy models that predict $c_{gw}\neq c$ at late cosmological times, put very strong bounds on the parameters of Horndeski theories, like NMDC \cite{Ezquiaga:2017ekz,Gong:2017kim}.
Due to the specific form of the GNMDC that we consider in this work, the non-GR term decouples at the end of inflation and issues related to the  
production of superluminal GWs at late times, become irrelevant{\footnote{Other viable subclasses of Horndeski theory should have a gravitational action of the conformal form, e.g a function $f(\phi)$ coupled to curvature \cite{Baker:2017hug, Creminelli:2017sry}.}.

Additionally to the new inflationary phenomenology, GNMDC might generate interesting features on the power spectrum of the primordial curvature perturbations, ${\cal P_R}(k)$, at scales smaller than the observed CMB scales.  The field dependence of the GNMDC term can modify the velocity of the inflaton field, influencing the amplitude of the curvature perturbations. If the amplification is strong enough, large primordial density perturbations can be generated, triggering primordial black hole (PBH) production. This possibility is very attractive, since most of the models found in the literature trigger PBH production through inflection point inflationary potentials \cite{Garcia-Bellido:2017mdw} or multi-field inflation; see e.g. for an early proposal \cite{Kawasaki:1997ju}.

In this work we take a step further, and utilize the possibilities of the GNMDC to construct single field inflationary models capable to generate PBHs. Non-canonical  kinetic terms and PBH production, e.g  from sound speed resonance during inflation has been examined also in \cite{Cai:2018tuh, Ballesteros:2018wlw, Chen:2019zza}.
Here we examine explicit GNMDC functions, $f(\phi)$, that dramatically decelerate the inflaton field at specific spots of the inflationary trajectory, amplifying the ${\cal P_R}(k)$. We estimate the values for the parameters of the models considered  so that a significant  abundance of PBHs can be produced at cosmologically interesting PBH mass windows. An explicit example with PBH mass  $M_\text{PBH}\sim 10^{21}$g is presented. 
The inflationary potential that we use is that of the Standard Model Higgs.

We note that a recent work \cite{Fu:2019ttf} discusses PBH production in a similar context. 
In our work, the GNMDC dominates during the entire stage of inflation and becomes fast negligible during the stage of oscillations.  Hence the inflationary predictions are purely due to the GNMDC dynamics, and we utilized the Higgs potential to describe and implement the PBH production. 
 
The paper is organized as follows. In Section \ref{setup} we give the basic set up of our theory and we derive the field equations. In Section \ref{sect} we calculate the power spectrum, the spectral index and the tensor-to-scalar ratio. In Section \ref{models} we discuss interesting and viable inflationary models in the framework of GNMDC, giving emphasis on the Standard Model Higgs inflation and inflation with exponential potentials.
In Section  \ref{SecObs} we determine the observational signatures of the GNMDC that distinguishes them from GR models.
In Section  \ref{SecPBH} we introduce the basics of PBH cosmology and we construct GNMDC models that trigger PBH production within Higgs inflation.
Finally, Section  \ref{SecConc} contains the conclusions of this work.


\section{The Setup - Derivation of the field equations }

\label{setup}

As we have discussed in the introduction the general term $G_5(\phi, X)$ which appears in the Horndeski Lagrangian   (\ref{lagran}) is well motivated and more specifically we will assume that it takes the  following  form
\begin{align} \label{GNMDC}
{\cal L}_5=G_5(\phi, X) \,G^{\mu\nu}\partial_\mu \phi \partial_\nu \phi\, =	\,f(\phi) \, \xi(X)\, G^{\mu\nu}\partial_\mu \phi \partial_\nu \phi \,,
\end{align}
where we take $\xi(X)=1$.
Our term introduces a  field dependent derivative coupling to the Einstein tensor. 
A motivated choice is $\alpha \phi^{\alpha-1} G^{\mu\nu}\ll M^{\alpha+1}$;
for $\alpha=0$ we get Einstein gravity and for $\alpha=1$ the simple NMDC is recovered. Choosing $\alpha>1$ modifies the phenomenology of the $\phi$ field-gravity system and is well motivated for two reasons. Firstly,
for $\alpha \phi^{\alpha-1} G^{\mu\nu}\gg M^{\alpha+1}$ (high friction limit - HF) the same friction effect as in NMDC appears, even though the specific inflationary predictions are expected to change.
Secondly, after the end of the inflationary stage, GR is expected to take over, since $\phi\rightarrow 0$ leads to $\alpha \phi^{\alpha-1} G^{\mu\nu}\ll M^{\alpha+1}$ at the end of inflation, switching off the GNMDC term. This would be a much desirable effect, because, as discussed in the introduction, the simple  NMDC case  sources  the late time instabilities and non-linearities.
In this work we show that term (\ref{GNMDC}) can yield a reheating period that is described  by GR gravity.

\subsection{The field equations}

Let us now proceed to the derivation of the equations that describe  the 
dynamics of the system. The  action of the theory we examine is

\begin{align} \label{gnmdcaction}
	S=\int d^4 x \sqrt{-g} \left [ \frac{M_\text{Pl}^2}{2} R-\frac{1}{2}\left(g^{\mu\nu}-
	f(\phi) G^{\mu\nu}\right)\partial_\mu \phi \partial_\nu \phi -V(\phi)\right] \,.
\end{align}
By varying with respect to the metric, the field equations are found,

\begin{align} \label{fieldEq}
 G_{\mu\nu} =\frac{1}{M^2_\text{Pl}} \left[
T_{\mu\nu}^{(0)} -f(\phi) \,T_{\mu\nu}^{(1)}-\frac{1}{2} f'(\phi) \, T_{\mu\nu}^{(2)}
 \right]~,
\end{align}
where $f'(\phi)=df/d\phi$ and $G_{\mu\nu}$ is the Einstein tensor.
The $T_{\mu\nu}^{(1)}$, $T_{\mu\nu}^{(2)}$ are given by
\begin{align}
T_{\mu\nu}^{(0)}&= \nabla_\mu\phi\nabla_\nu\phi  -  g_{\mu\nu} \left( \frac{1}{2}
 (\nabla \phi)^2 +V(\phi)
\right)~,  \\
	T_{\mu\nu}^{(1)}&=-G_{\mu\nu}\nabla_\lambda\phi \nabla^\lambda\phi+4 R^{\lambda}_{\text{ }\text{ }(\mu}\nabla_{\nu)}\phi\nabla_\lambda\phi-\nabla_\mu\phi\nabla_\nu\phi R+2 [\nabla^\kappa\phi \nabla^\lambda\phi R_{\mu\kappa\nu\lambda}+\nabla_\mu\nabla^\lambda\phi\nabla_\nu\nabla_\lambda\phi-\nabla_\nu\nabla_\mu\phi\nabla^2\phi]\nonumber \\
	&+g_{\mu\nu}[\nabla^2\phi\nabla^2\phi-\nabla_\kappa\nabla_\lambda\phi\nabla^\kappa\nabla^\lambda\phi-2 R_{\kappa\lambda}\nabla^\kappa\phi\nabla^\lambda\phi]~,\\
	T_{\mu\nu}^{(2)}&=g_{\mu\nu}(\nabla_\lambda\phi\nabla^\lambda\phi \nabla^2\phi-\nabla^\kappa\phi\nabla^\lambda\phi\nabla_\kappa\nabla_\lambda\phi)
+2\nabla^\lambda\phi\nabla_{(\mu}\phi\nabla_{\nu)}\nabla_\lambda\phi
-\nabla_\lambda\phi\nabla^\lambda\phi\nabla_\nu\nabla_\mu\phi-\nabla_\mu\phi\nabla_\nu\phi\nabla^2\phi~.
\end{align}
The parentheses enclosing indices stands for a due symmetrization on them\footnote{$A_{(\mu\nu)}=\frac{1}{2}(A_{\mu\nu}+A_{\nu\mu})$}. Notice also, that for $\alpha=1$ the term involving $T_{\mu\nu}^{(2)}$ switches off and one retrieves the equations of the usual NMDC.

\subsection{Friedmann Equations and the Klein-Gordon equation for the flat FLRW metric}

We assume a flat FLRW geometry, with $a(t)$ the scale factor.
 For a  homogeneous scalar field $\phi=\phi(t)$ the time-time and space-space components of the field equations (\ref{fieldEq}) read respectively,
\begin{align} \label{gnmdcFried00}
	3 M_\text{Pl}^2 H^2=V\left(\phi\right)+\frac{1}{2}\dot{\phi}^2+\frac{9}{2}\,f(\phi) \, \dot{\phi}^2 H^2~,
\end{align}
\begin{align} \label{gnmdcFried11}
	M_\text{Pl}^2 \left( \dot{H}+\frac{3}{2} H^2 \right)\, =\, \frac{V(\phi)}{2}-\frac{\dot{\phi}^2}{4}+f(\phi) \left[ \left( \frac{1}{2}\dot{H}+\frac{3}{4}H^2\right) \dot{\phi}^2+H \dot{\phi} \, \ddot{\phi}\right]+\frac12 \,f'(\phi) H\dot{\phi}^3~,
\end{align}
where an overdot denotes differentiation with respect to time, and $H(t)={\dot{a}(t)}/{a(t)}$.
The above equations are a  modified version of the Einstein equations for a minimally coupled field.  They are recast in the familiar form $\rho=3H^2 M^2$, $\rho+3p=-6M^2_\text{Pl}(H^2+\dot{H})$ respectively after defining the energy density and the pressure  for the scalar field,
\begin{align} \label{rhop}
&\rho_\phi \equiv \, \frac12 \dot{\phi}^2+ V(\phi) +\frac{9}{2} f(\phi) \dot{\phi}^2 H^2~, \\
& p_\phi \equiv \, \frac12 \dot{\phi}^2 - V(\phi) - \, f(\phi) \left[ \left(  \dot{H} +\frac32 H^2 \right) \dot{\phi}^2 +2 H\dot{\phi} \, \ddot{\phi}      \right]
-  f'(\phi)H \dot{\phi}^3~.
\end{align}

The Klein-Gordon equation derived by the action  \eqref{gnmdcaction} is
\begin{align}
\Bigl[\Bigl(\partial_\mu g^{\mu\nu} - & f(\phi)\,\partial_\mu G^{\mu\nu}\Bigr)
\partial_\nu\phi+
 \Bigl(g^{\mu\nu}-f(\phi)G^{\mu\nu}\Bigr)\partial_\mu\partial_\nu\phi-\frac12 f'(\phi)\, G^{\mu\nu}\partial_\mu\phi\partial_\nu\phi-\frac{dV}{d\phi}\Bigr]\sqrt{-g}\nonumber \\
	&-\frac{1}{2\sqrt{-g}}\Bigl[g^{\mu\nu}-f(\phi)G^{\mu\nu}\Bigr]\partial_\nu\phi\partial_\mu g =0~.
\end{align}
For $\phi=\phi(t)$, in a flat FLRW geometry, it becomes
\begin{align} \label{gnmdcKGeq}
	\ddot{\phi}\left (1+3f(\phi)H^2\right)
	+
	3H\dot{\phi} \left(1+3\, f(\phi)\,H^2+2\,f(\phi)\, \dot{H}\right)+\frac{3}{2}\, f'(\phi)\,\dot{\phi}^2H^2+\frac{dV}{d\phi}=0~.
\end{align}

\subsection{Slow Roll Approximation Parameters}

The first Hubble-flow function  in a $\phi$-dominated universe reads,
\begin{align} \label{Gsr1}
	\epsilon\equiv-\frac{\dot{H}}{H^2} & = \frac32 \frac{p+\rho}{\rho} \\
	&=\frac{3}{2} \frac{\dot{\phi}^2+ 3 f(\phi) \dot{\phi}^2 H^2 -  f(\phi) \left(   \dot{H}  \dot{\phi}^2 +2 H\dot{\phi} \, \ddot{\phi}      \right)
-  f'(\phi)H \dot{\phi}^3}{\rho_\phi}~.
\end{align}
Parameter $\epsilon$ is also called the first slow-roll parameter, though the above definition is general.
 We additionally define the slow-roll parameter,
\begin{align}
\delta \equiv \frac{\ddot{\phi}}{H\dot{\phi}}\,.
\end{align}
Slow-roll inflation  is realized when $\epsilon \ll 1$ and $\delta \ll 1$,  that is  $\dot{H} \ll H^2$ and $\ddot{\phi}\ll 3H\dot{\phi}$, hence the Friedmann and Klein-Gordon equations are approximated as 
\begin{align} \label{gnmdcFried00SRA}
	3 M_\text{Pl}^2 H^2\approx V(\phi)~,
\end{align}
\begin{align} \label{gnmdcKGeqSRA}
	3H\dot{\phi} \left(1 +3 \,f(\phi)H^2+ \frac12  f'(\phi)\, H \, \dot{\phi} \right) +\frac{dV}{d\phi}\approx 0\,.
\end{align}
Accordingly, the first slow-roll parameter is approximated by the expression
\begin{align} \label{SRap1}
\epsilon \simeq  \frac{3}{2}  \frac{\dot{\phi}^2}{\rho_\phi} \left(1+3H^2 f(\phi) - f'(\phi) H \dot{\phi}\right) \equiv \epsilon_\text{GR}+\epsilon_D+\epsilon_{\cal{B}}~.
\end{align}
We defined $\epsilon_\text{GR}=3\dot{\phi}^2/(2\rho_\phi)$, which corresponds to the first slow-roll parameter in GR gravity, $\epsilon_{\cal B}$  a new term proportional to $f'(\phi)$, and
\begin{align} \label{eD}
\epsilon_D\equiv \frac{3}{2} \frac{f(\phi) \dot{\phi}^2}{M^2_\text{Pl}}\,.
\end{align}
The effect of the GNMDC is that  the velocity $\dot{\phi}$ decreases $3H^2 f(\phi)$ times
and the overall result is that the $\epsilon$ significantly decreases. Indeed, in the slow-roll approximation, equation (\ref{SRap1})
is rewritten
\begin{align} \label{epsilon}
\epsilon \simeq \epsilon_V \, \frac{1+3H^2 f(\phi)-f'(\phi) H\dot{\phi}}{\left(1+3H^2 f(\phi) +f'(\phi)\,H \dot{\phi}/2 \right)^2}  \,  \equiv \,
\epsilon_V \, \frac{{\cal A}- {\cal B}}{\left({\cal A}+{\cal B}/2\right)^2} \,,
\end{align}
where $\epsilon_V=\frac{M_\text{Pl}^2}{2}\left(\frac{V'}{V}\right)^2$ and we defined
\begin{align}
&{\cal A } \equiv1+3H^2f(\phi)~,  \\
&{\cal B} \equiv f'(\phi)H\dot{\phi}~.
\end{align}
In the NMDC case it is $f'(\phi)=0$ and the first slow-roll parameter reads $\epsilon=\epsilon_V/{\cal A}$, hence the known result \cite{Tsujikawa:2012mk} is recovered.
On the same footing,  the second slow roll parameter $\eta$ is defined as $\eta \equiv \eta_V/ {\cal A}$,
where $\eta_V=M_\text{Pl}^2\frac{V''}{V}$.

\subsection{The number of $e$-folds }

Turning to the duration of the inflationary phase, the number of $e$-folds that take place from the initial moment $t$ until the end of inflation $t_\text{end}$ are
\begin{align}
N \equiv \int_t^{t_\text{end}} H dt=  \int^\phi_{\phi_\text{end}} \frac{H}{\dot{\phi}} d\phi \simeq    \frac{1}{M_\text{Pl}}\int^\phi_{\phi_\text{end}} \frac{{\cal A}+{\cal B}/2}{\sqrt{2\epsilon_V}} d\phi~,
\end{align}
where in the right hand side we considered the slow-roll approximation,  Eq. (\ref{gnmdcKGeqSRA}),
\begin{align}
\dot{\phi}= -\frac{V'(\phi)}{3H({\cal A}+{\cal B}/2)}\,.
\end{align}

\section{Power Spectrum, Spectral Index and the Tensor to Scalar Ratio}
\label{sect}

Each inflationary model predicts a spectrum of scalar and tensor perturbations, that makes the test against observations possible.
GNMDC dynamics affect the evolution of the universe when the inflaton field $\phi$  dominates the energy density. 
To study the perturbations, one can choose the gauge $\delta\phi=0$ to derive the power spectrum formula. 
In this particular gauge, any extra contributions to the perturbations due to the term
$f(\phi)\,G^{\mu\nu}\partial_\mu \phi \partial_\nu \phi$
are rendered equal to zero.

Following Refs. \cite{Tsujikawa:2012mk, Ema:2015oaa} the quadratic action  for the curvature perturbation ${\cal R}$ in the comoving 
gauge takes the form,
\begin{equation} \label{S2}
S_{(2)}\, = \, \frac{M^2_\text{Pl}}{2} \, \int dx^4 a^3 Q_s \left[ \dot{{\cal R}}^2 -\frac{c^2_s}{a^2} (\partial_i {\cal R})^2\right]~,
\end{equation}
where $Q_s \equiv {F^2 \,G}/({f(\phi) H^2})$ with
\begin{align}
F\equiv \frac{1-\epsilon_D/3}{1-\epsilon_D}~, \quad\quad
G\equiv\frac{\epsilon_D}{3}\left( 1+3H^2f(\phi)\frac{1+\epsilon_D}{1-\epsilon_D/3} \right)\,,
\end{align}
and $c^2_s$ is the sound speed squared, given by the expression
\begin{align} \label{cs2}
c^2_s=\left({1-\frac{\epsilon_D}{3}}\right)^{-1}\frac{\epsilon_D}{3 G}
\left[\left({1+{\epsilon_D}}\right) +3H^2f(\phi)\left[  \left(1+ {\epsilon_D}\right) +\frac{4}{9 F} \epsilon_D\right]  +6{\dot{H}} f(\phi) \left({1-\frac{\epsilon_D}{3}}\right)  \right]~.
\end{align}
For $f(\phi)\rightarrow 0$, we get $Q_s=\dot{\phi}^2/(2\,H^2 M^2_\text{Pl})$,  $c_s^2 \rightarrow 1$ and the canonical case is restored.
As we will discuss in the next section
the sound speed squared is found to be negative for some period during the oscillation in the NMDC case. In the framework of GNMDC this problem gets significantly ameliorated or even solved for appropriate values of the GNMDC parameters.

The power spectrum of the comoving curvature perturbation is
given by the expression
\begin{align} \label{PS-SR}
{\cal P}_{\cal R}=\frac{H^2}{8 \pi^2 \, Q_s\, c^3_s} \,.
\end{align}
During inflation, in the high friction (HF) limit, the condition $\epsilon_D \ll 1$ holds, hence
$Q_s \simeq \epsilon \, {\cal A}/({\cal A}-{\cal B})$ and $c_s\simeq 1$ and the power spectrum rewrites as:

\begin{align} \label{PR}
{\cal P}_{\cal R}=\frac{H^2}{8 \pi^2 \, M^2_\text{Pl}\, \epsilon_V}
\left({\cal A} + {\cal B}  +{\cal O}\left(\frac{{\cal B}^2}{{\cal A}}\right)  \right)~.
\end{align}
The change of the logarithm of the power spectrum per logarithmic interval $k$ gives the tilt of the scalar power spectrum,
\begin{align}
\frac{d \ln {\cal P}_{\cal R}}{{d \ln k}} \simeq -\left({\cal A} +\frac{{\cal B}}{2} \right)^{-1} \left[ 6\epsilon_V -2 \eta_V +\frac{V'}{V} M^2_\text{Pl} \frac{d}{d\phi} \ln\left({\cal A} +{\cal B} +{\cal O}\left(\frac{{\cal B}^2}{{\cal A}}\right) \right)\right]\,.
\end{align}
Assuming that ${\cal B}\ll {\cal A}$ and in the high friction limit of the slow-roll approximation we obtain
\begin{align}
\frac{d \ln {\cal P}_{\cal R}}{{d \ln k}}  \simeq -\frac{1}{\cal A} \left[ 8\epsilon_V -2 \eta_V +\frac{V'}{V} M^2_\text{Pl} \frac{f'(\phi)}{f(\phi)}\right]~. \\
\end{align}
Hence the spectral index, in the slow-roll approximation and in the HF limit, reads
\begin{align} \label{ns}
1-n_s \equiv -\frac{d\ln {\cal P}_{\cal R}}{d \ln k}\Bigr|_{k=aH}
 \simeq 8\epsilon - 2\eta + \epsilon  \,M_\text{Pl}\frac{f'(\phi)}{f(\phi)} \sqrt{\frac{2}{\epsilon_V}}~.
\end{align}
We see that the last term might turn the spectral index from red into blue.
We take advantage of this possibility later, where we examine PBH production due to the GNMDC.
Obviously, for $f'(\phi)=0$, i.e. the NMDC case with $f(\phi)=1/M^2$,
we obtain the conventional expression $1-n_s=8\epsilon-2\eta$.

The intrinsic tensor perturbation is decomposed into two independent polarization modes
\begin{equation}
S^{(2)}_t \, = \,\sum_p \int  \,  dx^4 a^3 Q_t \left[ \dot{h}^2_p -\frac{c^2_t}{a^2} (\partial h_p)^2\right]~,
\end{equation}
where $Q_t=M^2_\text{Pl}(1-\epsilon_D/3)/4$ and $c^2_t \simeq 1+2\epsilon_D/3$.  The tensor power spectrum is given by
\begin{align}
{\cal P}_t =\frac{H^2}{2\pi^2 Q_t c^3_t} \, \equiv  \, r \,{\cal P_R}\,.
\end{align}
Again, in the HF limit, it is  $\epsilon_D \ll 1$, hence the tensor-to-scalar ratio is given by the expression
\begin{align}
r= 16 \,\frac{\epsilon_V}{{\cal A}+{\cal B}}~.
\end{align}
Thus $r$ is smaller than in the simple NMDC case and also smaller than in the GR case.

\section{Towards viable inflation with  GNMDC}
\label{models}

From Eq. (\ref{gnmdcFried00}) we see that the Friedmann equation is rewritten as $3 H^2 M^2_\text{Pl}=(V+\dot{\phi}^2/2)(1-\epsilon_D)^{-1}$. A positive definite potential implies that $\epsilon_D < 1$.
Thus, the functions $F$ and $G$ are positive.
In order to avoid the appearance of scalar ghosts and Laplacian instabilities, we require that  $Q_s>0$ and $c^2_s >0$.
What is critical in our case is that $\dot{H}$ may turn from negative to positive during the oscillatory stage. Indeed, it is
\begin{align}
\dot{H}=- \epsilon \, H^2 =- \frac{p+\rho}{2M^2_\text{Pl}}~,
\end{align}
where $p$ is given by Eq. (\ref{rhop}). Rewriting it giving emphasis on the sign, we get
\begin{align}
 p_\phi = \, \left\{\frac12 \dot{\phi}^2 - V(\phi) \right\} - \,\dot{\phi}\, f(\phi)\left\{ \left[ \left(  \dot{H} +\frac32 H^2 \right) \dot{\phi} +2 H \, \ddot{\phi}      \right]
-  f'(\phi)H \dot{\phi}^2\right\}\, .
\end{align}
The first term in the brackets corresponds to the standard GR dynamics' pressure, while the second term is due to the GNMDC.

The problematic behavior appears because parameter $\epsilon$ might change sign during oscillations. 
We note that during oscillations, where the energy transforms from potential to kinetic energy in each period, the GNMDC becomes maximally effective when the kinetic energy dominates.
Hence the problem can be treated if the product $\dot{\phi} f(\phi)$ is smaller than the GR term $ \dot{\phi}^2$. 
This is achieved e.g by the choice $f(\phi)\sim 0$ whenever $\dot{\phi}$ takes the maximum value. 
Therefore it would be desirable if the coupling $f(\phi)$  vanished at the bottom of the inflaton potential.
Some working examples that we suggest have a GNMDC of the form
\begin{align} \label{phi-term}
f(\phi)=\frac{\alpha \phi^{\alpha-1}}{M^{\alpha+1}}\,~, \quad\quad \text{or} \quad\quad f(\phi)=\frac{1}{M^2}e^{\tau \phi/M_\text{Pl}}~.
\end{align}
For $\alpha=0$ (or $M\rightarrow \infty$) the GNMDC  is turned off, while for $\alpha=1$ (and $\tau=0$ for the exponential) the conventional NMDC coupling, which might be problematic during the reheating stage\footnote{Particular NMDC  models such as the Higgs inflation are found to be free from reheating instabilities   \cite{Germani:2015plv}.}, is recovered.  
We do not perform a full perturbative analysis for each model and we do not conclude whether gradient or phantom instabilities are completely evaded. 
We follow a simple graphic examination that gives suggestive results.
In Fig. \ref{fig:reheatplots} we examine the postinflationary behavior of the GNMDC models  comparing the canonical and non-canonical kinetic terms, and in Fig. \ref{fig:sspdcomparison} we depict the evolution of the sound speed squared.

\begin{figure}[htbp]
\makebox[\textwidth]{
\includegraphics[width=0.45\textwidth]{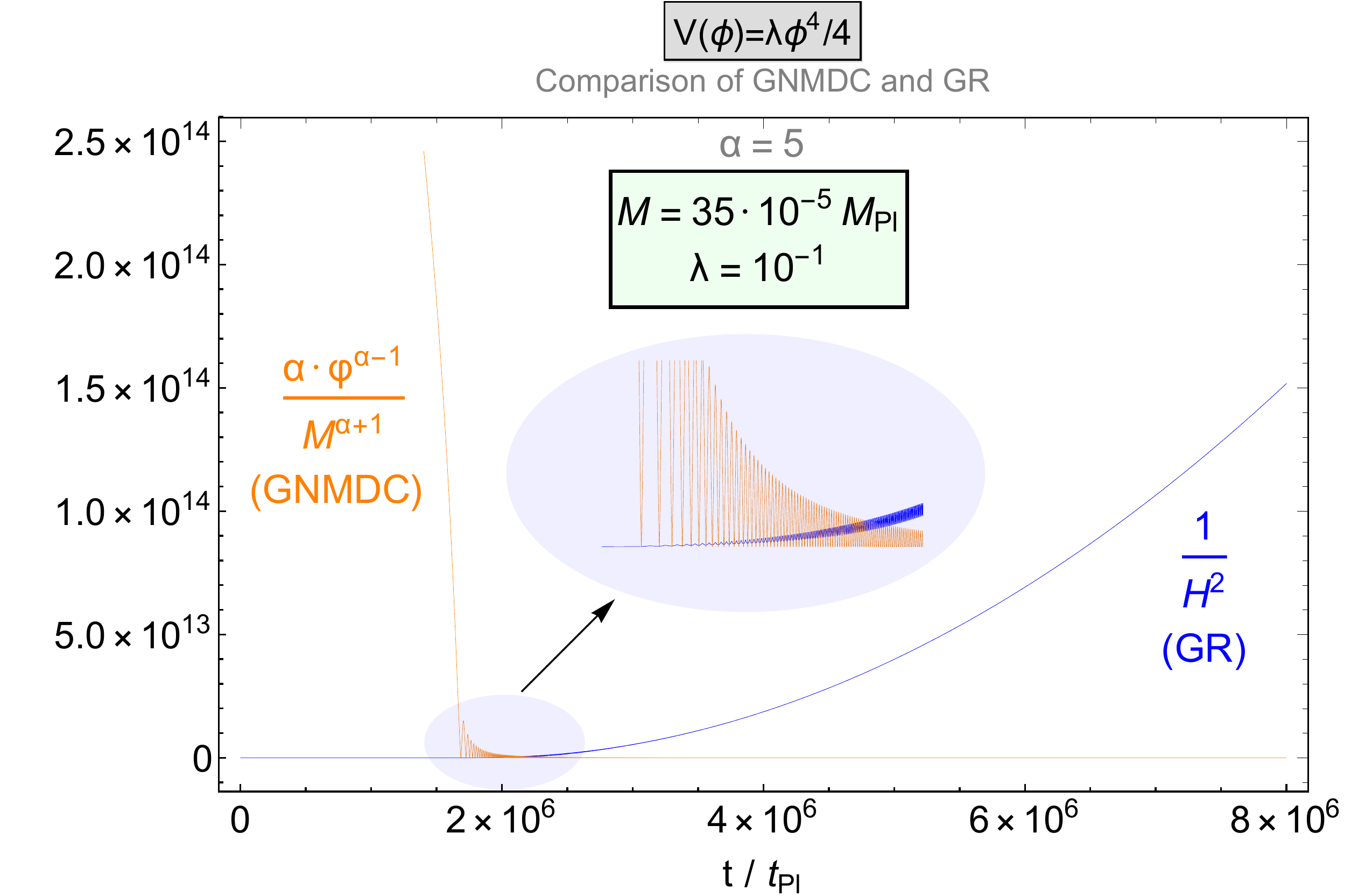}%
\hfill
\includegraphics[width=0.45\textwidth]{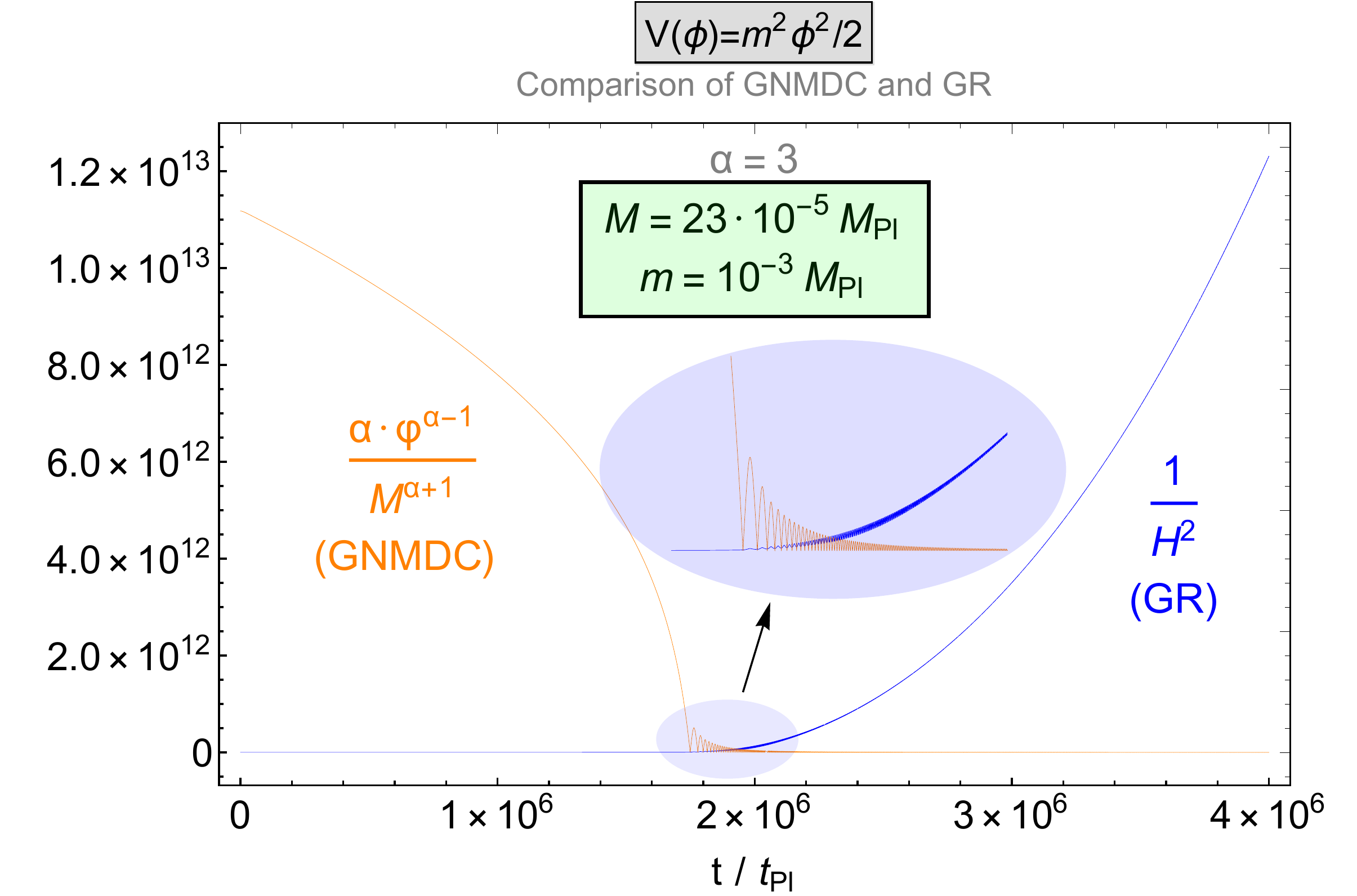}%
}\\[0.4cm]

\caption{These plots depict the decay of the GNMDC term after the end of inflation for a Higgs ({\it left panel}) and a quadratic ({\it right panel}) potential.
The GR term  takes over the GNMDC term after a few oscillations depending on the value of $\alpha$. In contrast the  NMDC ($\alpha=1$) term  dominates for a vastly longer period leading to potentially problematic  instabilities. The verical axis has $1/M_\text{Pl}^2$ units.}
\label{fig:reheatplots}
\end{figure}

\begin{figure}[htbp]
\makebox[\textwidth]{
\includegraphics[width=0.6\textwidth]{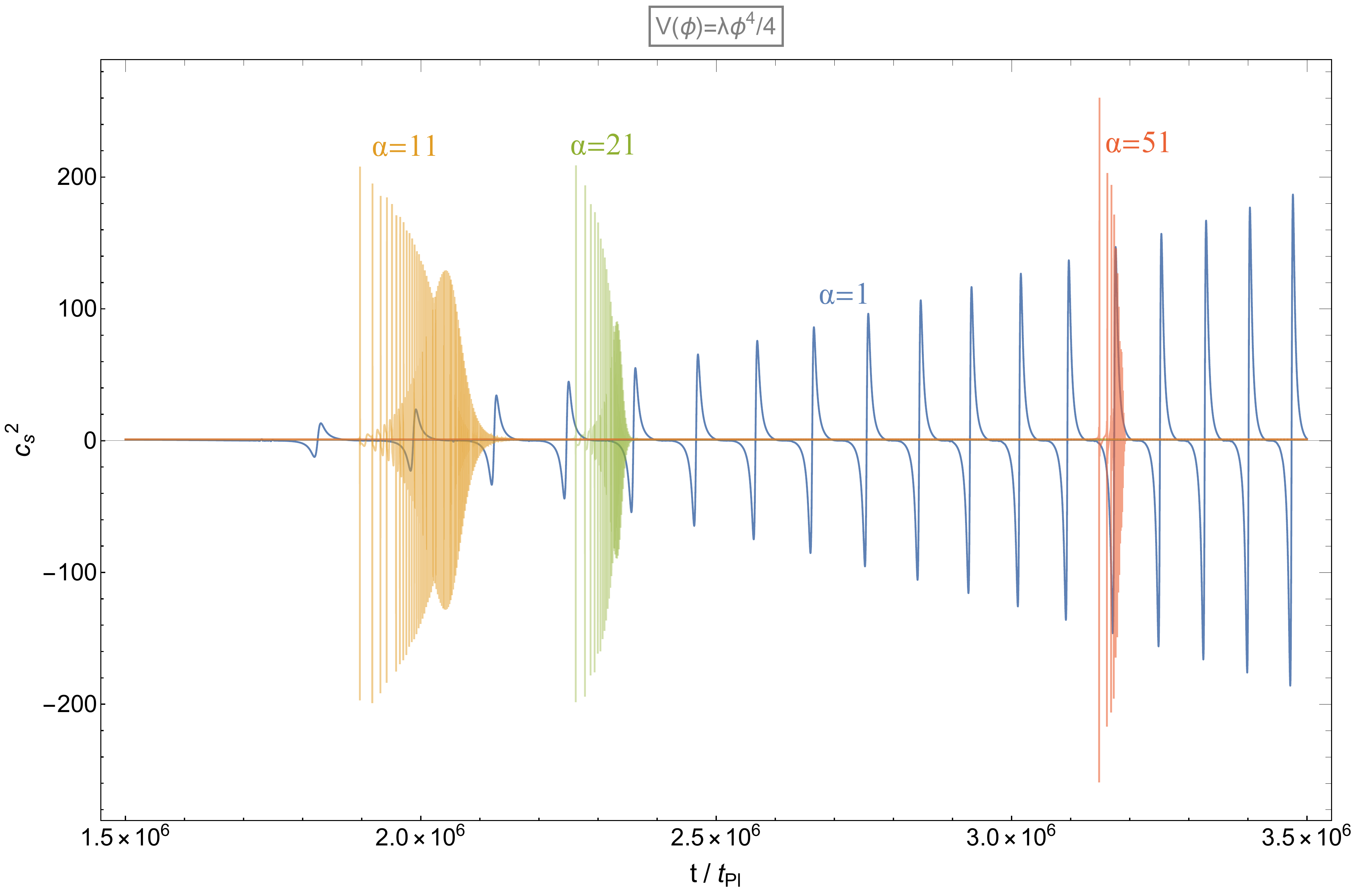}%
}\\[0.4cm]

\caption{The comparison of the $c_s^2$ for the Higgs potential, for various cases of the parameter $\alpha$ of the GNMD-Coupling. All the examples yield 60 e-folds, and are $\mathcal{P}_\mathcal{R}$-normalized. For $\alpha=1$ (NMDC) the oscillations increase and do not decay, causing potential post-inflationary instabilities. This behavior is due to the fact the NMDC remains dominant with respect to GR after inflation. However, as $\alpha$ increases, the GNMDC decays faster after inflation.  
This leads to $c_s^2$ becoming equal to one  after  a distinct number of oscillations and GR takes over, ameliorating the problem of post-inflationary instabilities compared to the simple NMDC case.}
\label{fig:sspdcomparison}
\end{figure}

\subsection{Inflationary observables}

The main inflationary observables are the amplitude of the scalar perturbations ${\cal P_R}(\phi_\text{cmb})$, the scalar tilt $n_s$, and the tensor-to-scalar ratio $r$.
In the HF limit it is ${\cal A} \gg 1$, implying ${\cal B}\sim f'(\phi)/{\cal A} \ll {\cal A}$, hence the power spectrum reads approximately
\begin{align}\label{PR2}
\left. {\cal P}_{\cal R}(\phi, \lambda_p, \alpha, M) \right|_{\phi=\phi_\text{cmb}} \, \simeq \frac{H^2}{8 \pi^2 \, M^2_\text{Pl}\, \epsilon_V} {\cal A}\, \simeq \,
\frac{V^2(\phi)}{24 \pi^2 \, M^6_\text{Pl}\, \epsilon_V(\phi)}  f(\phi, \alpha, M)\,= \, 2.2 \times 10^{-9} \,.
\end{align}
The above equation yields a constraint for the parameters of the GNMDC  $f(\phi, M,...)$  function  in relation with the potential parameter
at ${\phi=\phi_\text{CMB}}$. $\phi_\text{CMB}$ is the field value at the moment that the CMB scale $k_\text{cmb}=0.05$ Mpc$^{-1}$, probed by the Planck satellite \cite{Akrami:2018odb}, exited the Hubble horizon.  It is determined by the required number of $e$-folds 
 \begin{align} \label{N}
N \, \simeq \, \frac{1}{M^2_P} \, \int_{\phi_\text{end}}^{\phi} \frac{\epsilon_V}{\epsilon} \, \frac{V}{V'} \, d\phi \,.
\end{align}
The scalar tilt (\ref{ns}) reads
\begin{align}
1-n_s  \simeq \frac{M^2_\text{Pl}}{V(\phi, \lambda_p) f(\phi, \alpha, M)} \left( 8\epsilon_V -2\eta_V+M_\text{Pl} \frac{f'(\phi)}{f(\phi)}\sqrt{2\epsilon_V(\phi, \lambda_p)} \right) \,~.
\end{align}

 Although GNMDC inflation dynamics are multi-parametric, the above relations can  constrain the parameters of inflation  models.
Let us be specific and examine, in what follows, the dynamics of particular GNMDC terms and inflationary potentials.

\subsubsection{Inflation with power law monomial potentials and GNMDC}

In order to demonstrate the effects of the GNMDC  let us consider the general class of power law monomial potentials, with a derivative coupling of similar form
\begin{equation} \label{monomial}
V(\phi)= \lambda_p \,\phi^p\,, \quad\quad f(\phi)\, = \, \alpha \frac{\phi^{\alpha-1}}{M^{\alpha+1}}
\end{equation}
where $\lambda_p$ has dimensions of mass to the power $4-p$.
Later, we will  focus on the $\phi^4$  model, that is the Higgs potential, known to be realized in nature.
In order to examine the parameter space of this particular GNMDC term, we explicitly write the dynamic equation for the $\phi$ field, after eliminating  the $H(t)$, that reads
\begin{align}
	&\ddot{\phi}+V'(\phi)+\frac{2 \sqrt{3} \alpha  \phi^\alpha \dot{\phi} V(\phi) \sqrt{\frac{\phi  M^{\alpha +1} \left(\dot{\phi}^2+2 V(\phi )\right)}{2 M_\text{Pl}^2  M^{\alpha +1} \phi-3 \alpha  \phi^\alpha \dot{\phi}^2}}}{2 M_\text{Pl}^2  M^{\alpha +1} \phi-\alpha  \phi ^\alpha \dot{\phi}^2}+\sqrt{3} \dot{\phi} \sqrt{\frac{\phi  M^{\alpha +1} \left(\dot{\phi}^2+2 V(\phi )\right)}{2 M_\text{Pl}^2  M^{\alpha +1} \phi-3 \alpha  \phi ^{\alpha } \dot{\phi}^2}}\nonumber \\
	&+\frac{\alpha  \left(\dot{\phi}^2+2 V(\phi )\right) \left((\alpha -1) \dot{\phi}^2+2 \phi  \ddot{\phi}\right)}{\alpha  \phi  \dot{\phi}^2-2 M_\text{Pl}^2 M^{\alpha +1} \phi ^{2-\alpha }} \nonumber \\
	&=\frac{\sqrt{3} \alpha  \phi ^{\alpha } \dot{\phi}^3 \sqrt{\frac{\phi  M^{\alpha +1} \left(\dot{\phi}^2+2 V(\phi )\right)}{2 M_\text{Pl}^2  M^{\alpha +1} \phi-3 \alpha  \phi ^{\alpha } \dot{\phi}^2}}}{2 M_\text{Pl}^2  M^{\alpha +1} \phi-\alpha  \phi ^{\alpha } \dot{\phi}^2}+\frac{3 \alpha  \left(\dot{\phi}^2+2 V(\phi )\right) \left((\alpha -1) \dot{\phi}^2+2 \phi  \ddot{\phi}\right)}{6 \alpha  \phi  \dot{\phi}^2-4 M_\text{Pl}^2  M^{\alpha +1} \phi ^{2-\alpha }}~.
\end{align}
Asking for positive square roots and the absence of poles, we specify the allowed phase space
$\{\phi, \dot{\phi}\}$ and certain relations between the constants $M$ and $\alpha$ are obtained.

 For ${\cal A} \gg {\cal B}$
the slow-roll  parameters read
\begin{align}
\epsilon(\phi) \simeq \frac{M^2_\text{Pl}}{2\phi^2}\frac{p^2}{{\cal A}(\phi)} = \frac{1}{2} \eta~.
\end{align}
From Eq. (\ref{N}) the number of $e$-folds is found,
\begin{align}
N(\phi)\,\simeq \, \frac{\alpha \lambda_p}{p\, (p+a+1)} \, \frac{1}{M^4_\text{Pl}\, M^{\alpha+1}} \left( \phi^{p+\alpha+1}- \phi^{p+\alpha+1}_\text{end}\right)\,.
\end{align}
The end of inflation is determined by the solution of the equation $\epsilon=1$, i.e. $\epsilon_V \, ({\cal A} - {\cal B})/\left({\cal A}+{\cal B}/{2}\right)^2 = 1$,
that yields $\phi_\text{end} =\left(\frac{p^2}{2\,\alpha\, \lambda_p} \,M^4_\text{Pl}\, M^{\alpha+1} \right)^{1/(a+p+1)}$.
For $\phi^{p+\alpha+1} \gg \phi^{p+\alpha+1}_\text{end}$ it is
\begin{align} \label{phi-cmb}
\phi^{p+\alpha+1}(N)\,\simeq \, \frac{p(p+\alpha+1)}{\alpha \lambda_p} \, {M^4_\text{Pl}\, M^{\alpha+1}}\, N\,.
\end{align}
Hence, the slow-roll parameters written in terms of the $e$-folds number are,
\begin{align}
\epsilon \simeq
\frac{p}{2(p+\alpha+1)}\, \frac{1}{N} \,, \quad \eta= \frac{2p-2}{p} \epsilon\,.
\end{align}
Finally the scalar tilt, for $\alpha\geq 1$, is given by the expression
\begin{align}
1-n_s \simeq \, \frac{p\,(2p+ \alpha+1) }{\alpha\lambda_p}\frac{M^4_\text{Pl} M^{\alpha+1}}{\phi^{p+\alpha+1}} \, =\,  \frac{2(2p+\alpha+1)}{p}\, \epsilon~,
\end{align}
and the dependence of the spectral tilt of the scalar power spectrum and the tensor-to-scalar ratio on the $e$-folds number is obtained,
\begin{align} \label{ns-r}
1-n_s\,\simeq  8\epsilon -2\eta +2 \epsilon \frac{\alpha-1}{p}=
\frac{2p+\alpha+1}{p+\alpha+1}\, \frac{1}{N}\,, \quad r\simeq \frac{8 \, p}{p+\alpha+1 }\, \frac{1}{N}~,
\end{align}
for $\alpha\geq 1$.
Note that the $n_s$ and $r$ do not depend on the potential, $\lambda_p$ parameter, nor the GNMDC mass scale $M$.
For larger values of $\alpha$ $n_s$ increases and $r$ decreases; this behavior is depicted in the plots of Fig. 1, 2 and 4.
The requirement of $r<0.064$ and the fiducial spectral index value $n_s=0.965$
can be both satisfied for an $e$-folds number $N\lesssim 37$.
Therefore inflation models such as the quartic or the quadratic power-law inflation models
can be completely compatible with the Planck 2018 constraints \cite{{Akrami:2018odb}}, if proper functions for the GNMDC are chosen.

Summarizing, for a particular monomial inflationary potential with power $p$ the measured value for the spectral index gives a constraint on $N$ and $\alpha$, see Eq. (\ref{ns-r}).
  Eq. (\ref{phi-cmb}) gives the field value that corresponds to the CMB pivot scale, $\phi_\text{CMB}$, and, for that value, the CMB normalization constraint (\ref{PR2}) determines the GNMDC scale $M$ and parameter $\alpha$.
 If we assume a particular value for $N_\text{}$ (for example, for a particular postinflationary evolution, as we do in section \ref{PBHsec} where we discuss PBH generation) the $\alpha$ parameter can also be specified.
 Hence, a particular inflationary model, e.g. the Higgs potential that is realized in nature, can give viable GNMDC inflation for appropriate values of $M$ and $\alpha$, specified by observations. In addition  Higgs inflation with GNMDC is possible to be weakly coupled during the inflationary evolution on the same footing with the results of \cite{Germani:2014hqa}.

\begin{figure}[htbp]
\makebox[\textwidth]{
\includegraphics[width=0.46\textwidth]{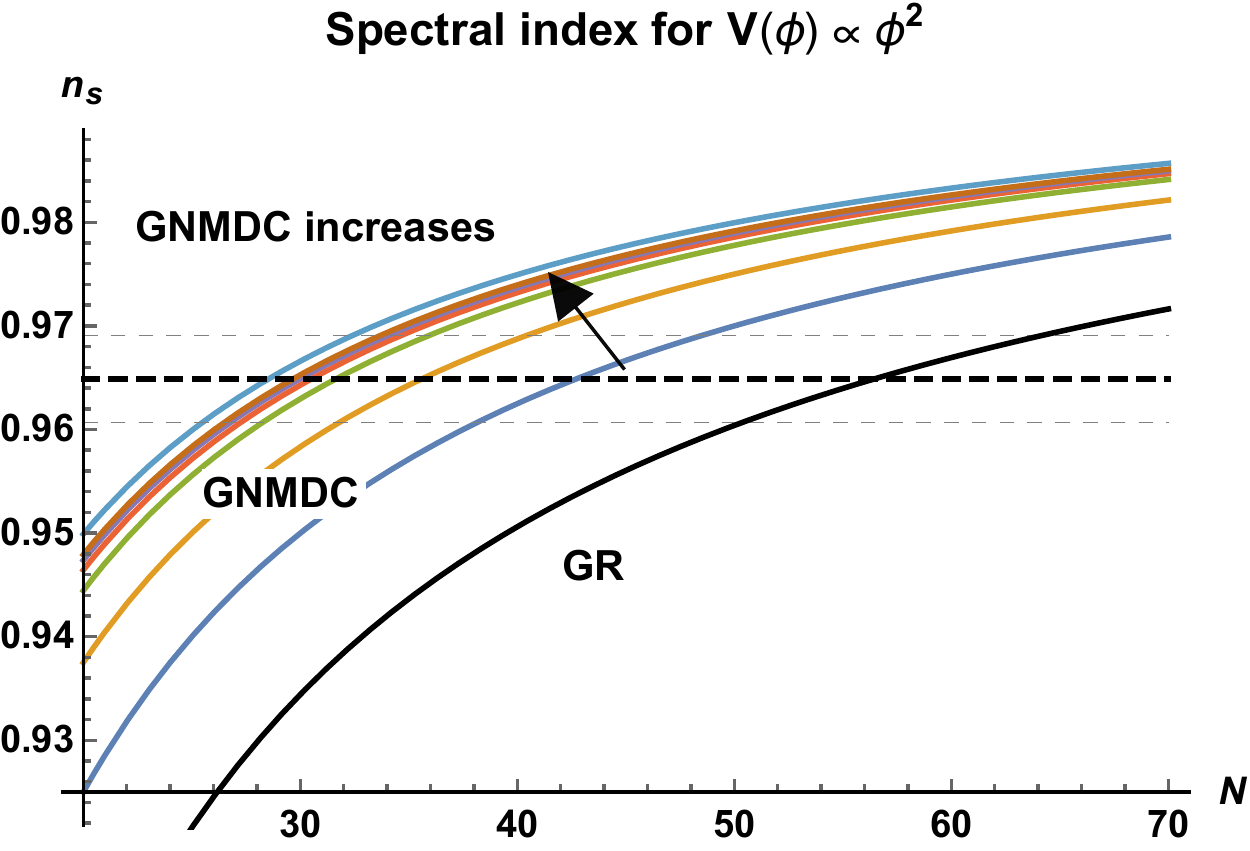}%
\quad 
\includegraphics[width=0.46\textwidth]{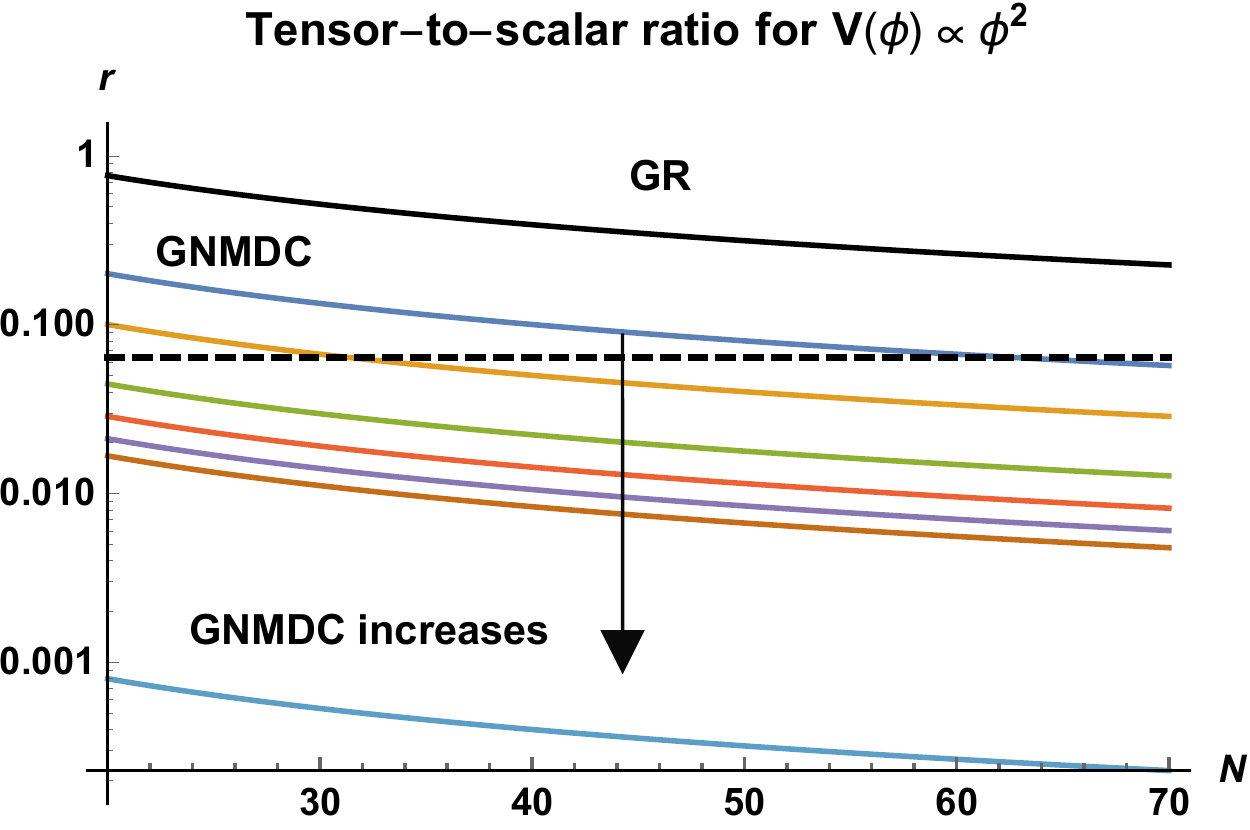}%
}\\[0.3cm]
\caption{
The inflationary predictions for a quadratic potential $V(\phi)=\lambda_2 \phi^2$ with GR and with GNMDC for $f(\phi)=\phi^{\alpha-1}/M^{\alpha+1}$ and $\alpha=1, \, 5, \, 15, \, 25,\, 35,\, 45$ and $1000$.
{\it Left panel}: The dependence of the spectral index on the $e$-folds number $N$, for different $\alpha$ values; $\alpha$ increases from bottom to top and the curves do not depend on $\lambda_2$.
The band enclosed by the dashed lines indicates the
Planck 2018  measured value for the spectral index at $68\%$ CL.
{\it Right panel}: The dependence of the tensor-to-scalar ratio on $N$ and $\alpha$.  Here, $\alpha$ increases from top to bottom.
Apparently, large values for $\alpha$ predict small values for $r$.
The dashed line is the upper limit on tensor perturbations set by the Planck 2018.
}
\label{figQuadratic}
\end{figure}

\begin{figure}[htbp]
\makebox[\textwidth]{
\includegraphics[width=0.46\textwidth]{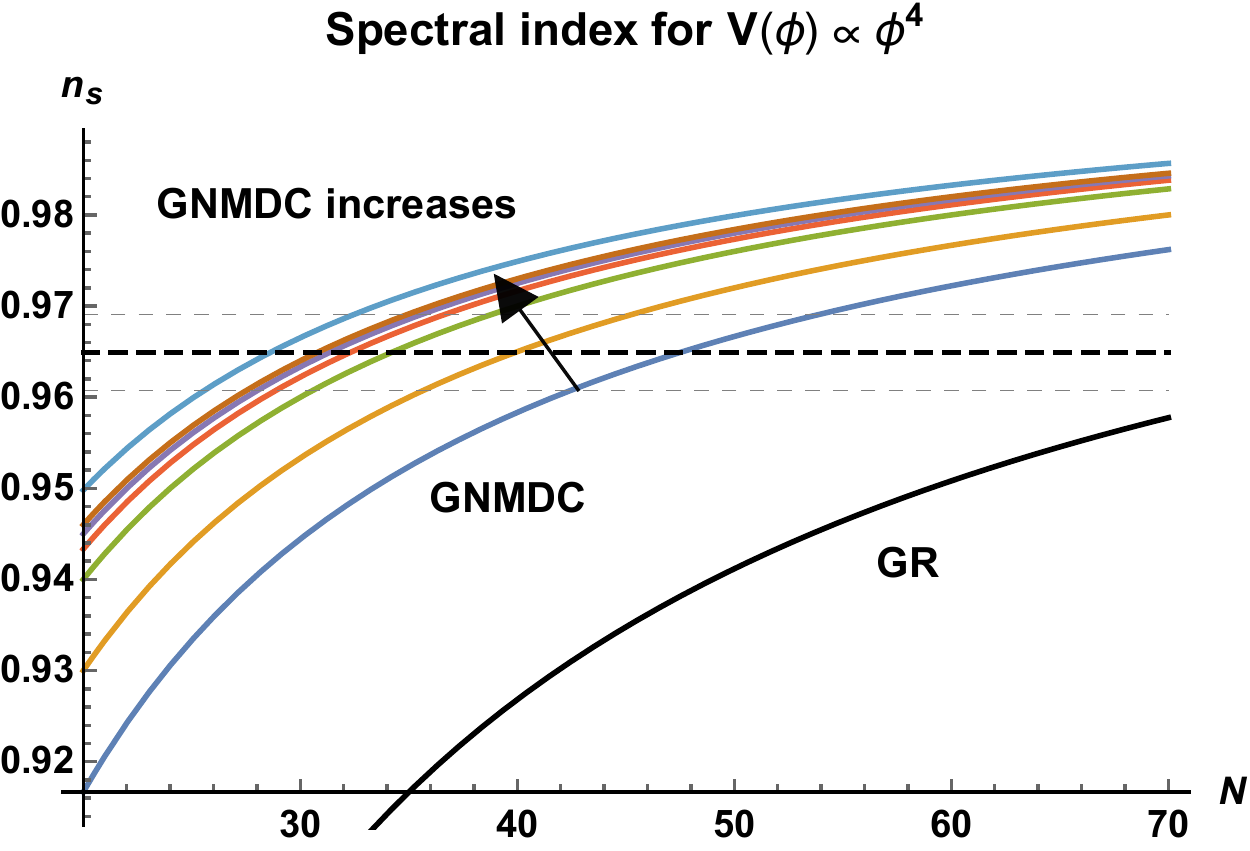}%
\quad 
\includegraphics[width=0.46\textwidth]{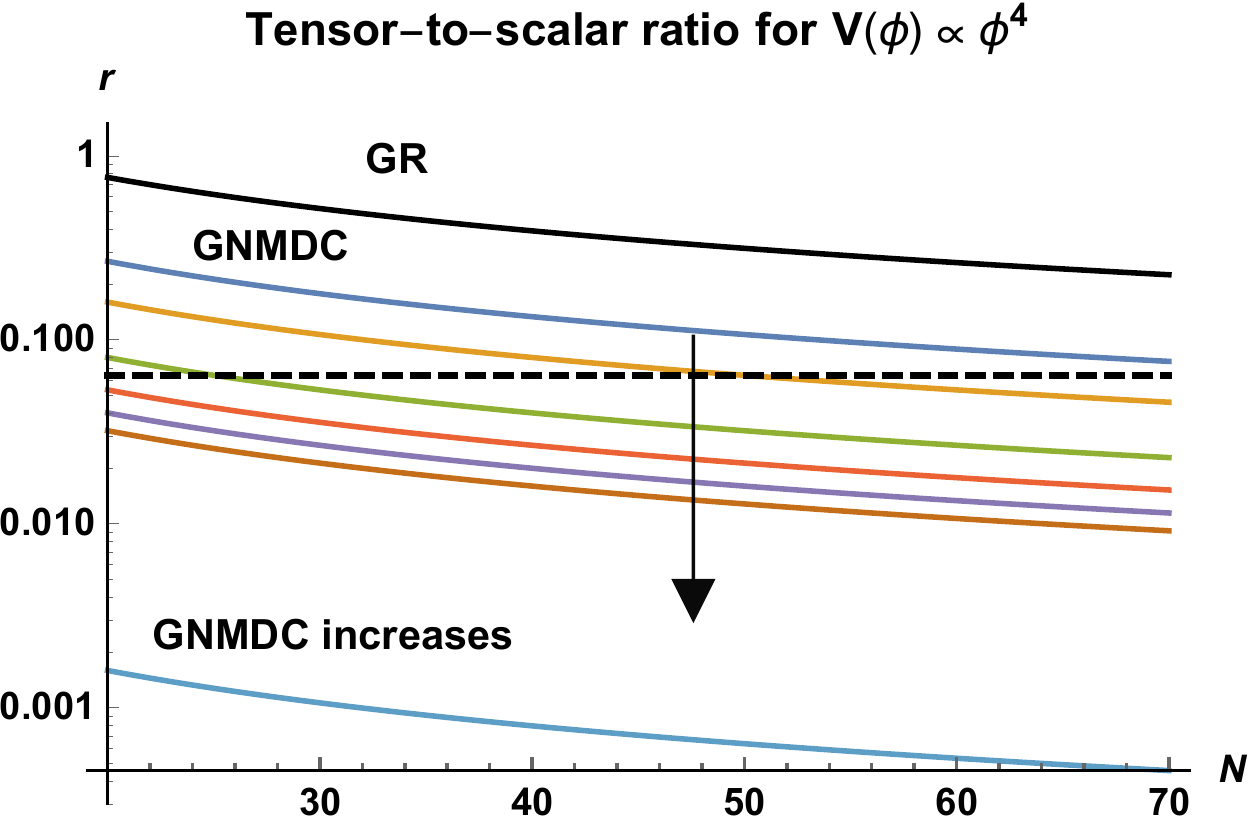}%
}\\[0.3cm]
\caption{As in Fig. \ref{figQuadratic}, for a Higgs-like potential, $V(\phi)=\lambda_4 \phi^4$.}
\label{figQuartic}
\end{figure}

To illustrate our results, we plot in Figs. \ref{figQuadratic} and \ref{figQuartic}
the $n_s(N)$ and $r(N)$ relations for various $\alpha$ values, stressing the different inflationary phenomenology with respect to the simple NMDC case and to the case that GR dominates.
Moreover, in Figs. \ref{fig:rofnsplots}-\ref{fig:examplehiggs} we show exact results as examples of GNMDC dynamics.
  Specifically, in Fig. \ref{fig:rofnsplots}  we plot the $r=r(n_s)$ contour curves for $\alpha=1,3,5,7$, for a  Higgs and a quadratic potential against the Planck 2018 $68\%$ and $95\%$ C.L. contour regions \cite{Akrami:2018odb}, for particular numbers of $e$-folds $N$.
In Figs. \ref{fig:examplequad} and \ref{fig:examplehiggs} we see examples of the oscillations of the inflaton, as well as its $\phi(N)$ graph for a quadratic potential with $\alpha=3$ and a Higgs potential with $\alpha=5$, with parameters specified in each  of the graphs. For reference, the evolution of the NMDC ($\alpha=1$) case with the same parameters is also shown, to highlight the fact that the GNMDC models achieve a slow-roll stage easier than the simple NMDC.

Moreover, it is important to examine the behavior of the GNMDC models in the post-inflation era. As shown in \cite{Ema:2015oaa, Germani:2015plv, Dalianis:2016wpu} the  postinflationary NMDC  dynamics made the inflaton field to oscillate vividly and possible instabilities  could plague the reheating stage and the NMDC  inflation models. To see the corresponding behavior of GNMDC, we compare  the canonical GR kinetic term and the GNMDC term near and after the end of inflation.  
Indeed, as we demonstrate in Fig. \ref{fig:reheatplots}, the non-canonical kinetic term evolves in such a way 
 so that the GR dynamics take over after a few oscillations  (depending on the value of $\alpha$),   a fact that can protect the inflation model from possible instabilities, contrary to the simple NMDC inflation models.

\subsubsection{Inflation with exponential potentials and GNMDC}

Exponential potentials are of particular interest in beyond the Standard Model set ups since they are connected to the superstring dilaton. Within the NMDC framework they were examined in \cite{Dalianis:2014nwa}. 
Let us here consider exponential potentials, with an exponential GNMDC,
\begin{equation} \label{expon}
V(\phi)=V_0 e^{2\lambda \phi/M_\text{Pl}}\,, \quad\quad f(\phi)\, = \,  \frac{e^{2\tau \phi/M_\text{Pl}}}{M^2}~.
\end{equation}
The slow-roll parameters are $\epsilon_V=2 \lambda^2$ and $\eta_V=2\epsilon_V$. For ${\cal A} \gg {\cal B}$ it is $\epsilon \simeq \epsilon_V/{\cal A}$ and
\begin{align}
\epsilon(\phi) \simeq \frac{2\lambda^2}{{\cal A}(\phi)} = \frac{1}{2} \eta(\phi)\,.
\end{align}
The condition ${\cal A} \gg {\cal B}$ is satisfied for $\lambda \tau \ll 3H^2 f(\phi)\simeq {\cal A}$. The ${\cal A}$ value is related with the power spectrum value (\ref{PR2}) and it is ${\cal A}=16 \epsilon_V/r$. Hence, roughly, for $\tau< \lambda/r$
it is ${\cal A} \gg {\cal B}$.

Interestingly enough, the slow-roll parameters do depend on the field value contrary to the GR case. This implies that inflation models with exponential potentials are not eternal in the GNMDC case and a graceful exit can be realized.
The scalar tilt Eq. (\ref{ns}) is given by the expression,  
\begin{align}
1-n_s
 \simeq 8\epsilon - 2\eta + \epsilon  \,M_\text{Pl}\frac{f'(\phi)}{f(\phi)} \sqrt{\frac{2}{\epsilon_V}} \simeq 4\epsilon+\epsilon \frac{2\tau}{\lambda}  \simeq 2\epsilon \left( 2+\frac{\tau}{\lambda}\right)~.
\end{align}
The number of $e$-folds is found to be
 \begin{align} \label{Nexp}
\left. N({\phi, \phi_{\text{end}}}) \simeq \frac{{\cal A}(\phi)}{4 \lambda(\lambda+\tau)} \right|_{\phi_{\text{end}}}^\phi \simeq \frac{\lambda}{2(\lambda+\tau)} \frac{1}{\epsilon(\phi)}~,
\end{align}
where in the last equality we considered that ${\cal A}(\phi) \gg {\cal A}(\phi_\text{end})=2\lambda^2$ and approximated $N(\phi, \phi_{\text{end}}) \simeq N(\phi)$. Inflation ends at the value,
\begin{align}
\phi_\text{end} =\frac{M_\text{Pl}}{2(\lambda+\tau)} \ln \left(\frac{2\lambda^2 M^2 M^2_\text{Pl}}{V_0} \right)~.
\end{align}
Therefore, the scalar tilt and the tensor-to-scalar ratio read, in terms of the $e$-folds and the parameters $\lambda$ and $\tau$,
\begin{align} \label{nsexp}
1-n_s \simeq \frac{2\lambda +\tau}{\lambda+\tau} \, \frac{1}{N}\,, \quad r\simeq \frac{8 \, \lambda}{\lambda+\tau }\, \frac{1}{N}~.
\end{align}
As $\tau$ increases the $1-n_s$ converges to the value $1/N$ while the tensor-to-scalar ratio decreases.
For the fiducial value $n_s=0.965$ \cite{Akrami:2018odb}, $\lambda$ and $\tau$ are both positive, which means that the GNMDC decreases towards the end of inflation, for  $ 28 \lesssim  N \lesssim 57$.
Moreover, if we ask for $r<0.064$ the ratio of the parameters $\lambda, \tau$ is bounded, $\lambda/\tau<(125/N-1)$.  The spectral index and the tensor-to scalar ratio constraints are both satisfied for $e$-folds number $N<47$.
In total the parameters $\lambda, \tau$ and $M$ can be specified from the CMB normalization, the $n_s$ value and the $e$-folds number $N$.

In our opinion,  the fact that the exponential potential augmented with GNMDC can well be in accordance with the Planck 2018 data \cite{Akrami:2018odb} is exciting.

\begin{figure}[htbp]
\makebox[\textwidth]{
\includegraphics[width=0.46\textwidth]{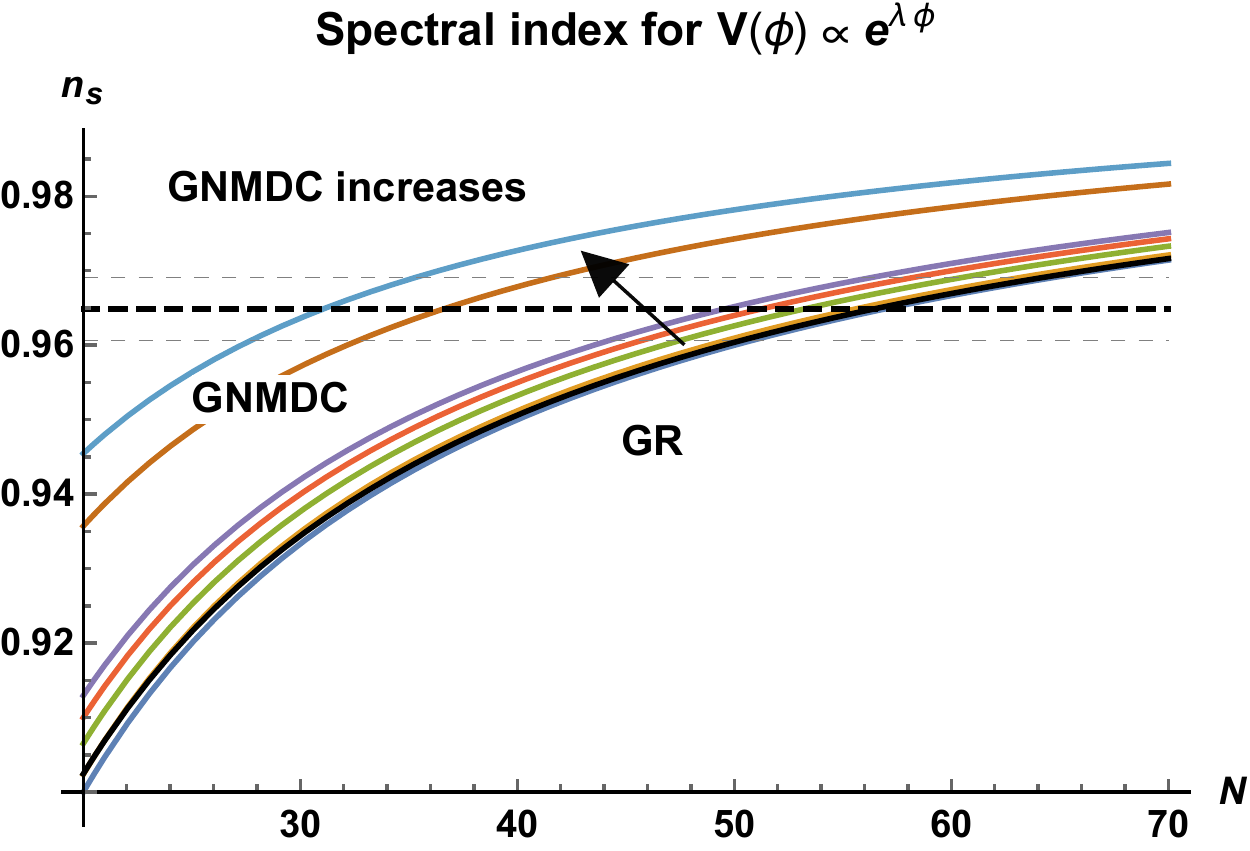}%
\quad 
\includegraphics[width=0.46\textwidth]{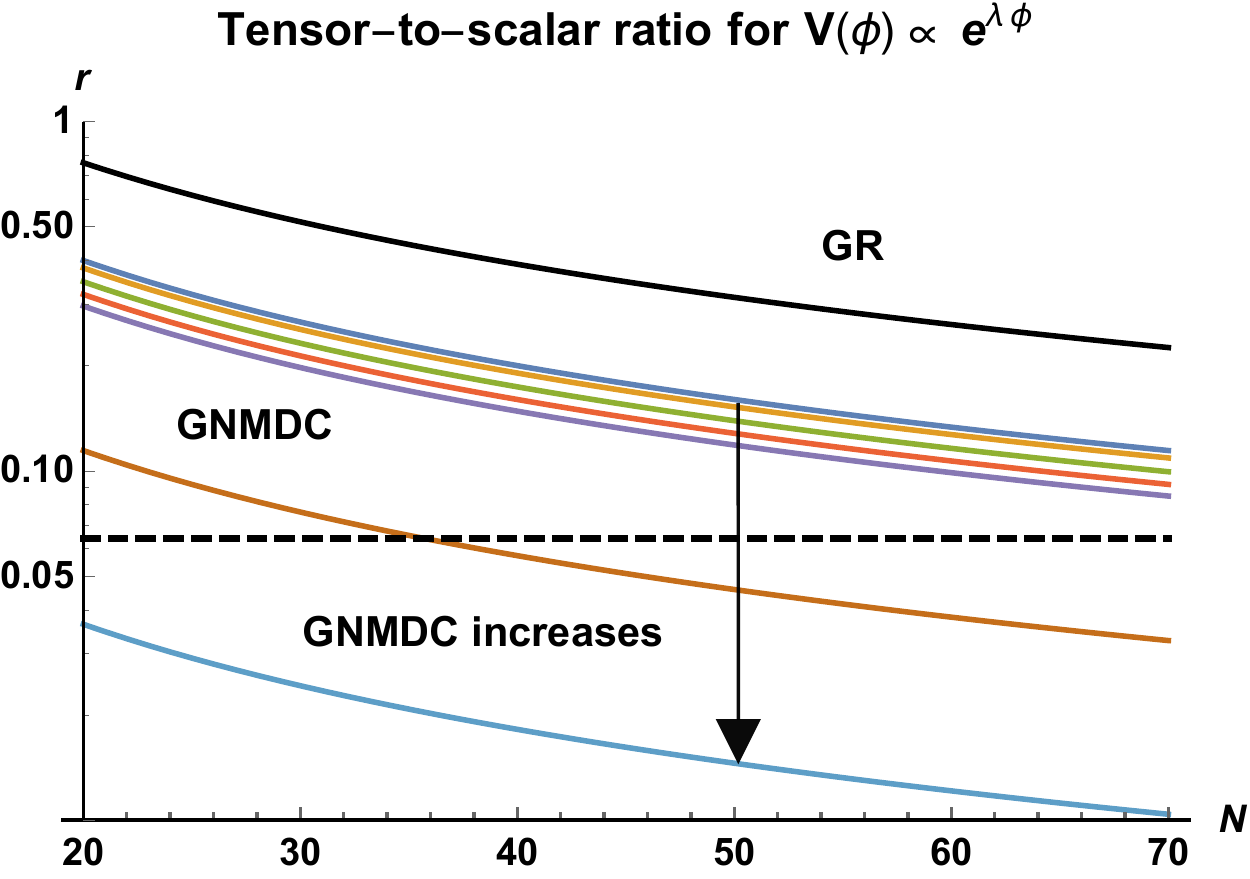}%
}\\[0.3cm]
\caption{The inflationary predictions for exponential potential $V(\phi)=V_0 e^{2\lambda \phi/M_\text{Pl}}$ within GR and with GNMDC function $f(\phi)=M^{-2} e^{2\tau\phi/M_\text{Pl}}$, for $\lambda=100$ and $\tau=1, \, 5, \, 15, \, 25,\, 35,\, 250$ and $1000$.
{\it Left panel}: The dependence of the spectral index on the $e$-folds number $N$ for different $\tau$ values; $\tau$ increases from bottom to top and the curves do not depend on $\lambda$.
The band enclosed by the dashed lines indicates the
Planck measured value for the spectral index at $68\%$ CL.
{\it Right panel}: The dependence of the tensor-to-scalar ratio on $N$ and $\tau$ values.  Here, the $\tau$ value increase from top to bottom.
Large values of $\tau$ predict small values for $r$.
The dashed line is the upper limit on tensor perturbations set by the Planck 2018.}
\label{figexp}
\end{figure}


\section{The observational signatures of the GNMDC}
\label{SecObs}

Relations (\ref{ns-r}) and (\ref{nsexp}) are the essential inflationary predictions of the GNMDC models examined in the previous section.
Due to the large number of inflationary models that exist in the literature
we will examine here whether these predictions are distinct, and if not how the possible degeneracy breaks.
We will see that during slow-roll, the predictions are similar to GR models with fractional power-law potentials.
This similarity,  owed to a correspondence between the GNMDC and GR dynamics, is also helpful to  tackle the predictions of the subtle GNMDC dynamics. Nevertheless, the similarity is not exact and any degeneracy breaks when both  the inflationary and postinflationary  evolution of the inflaton field is considered.

\subsection{Correspondence between GNMDC and GR models}
In the HF limit, 
where the GNMDC dominates, and during slow roll a very interesting and informative correspondence between GNMDC and GR dynamics can be found.
In our models the picture of the canonical scalar field that the GNMDC model corresponds to is clear during the HF period.
In this regime the approximate Friedmann and the Klein-Gordon equations read
\begin{align} \label{syst}
H^2\simeq \frac{V(\phi)}{3M^2_P}\,,\quad\quad 3H\dot{\phi} \simeq - \frac{\epsilon}{\epsilon_V}V'(\phi)\,,
\end{align}
where the ratio ${\epsilon_V}/{\epsilon}$ is given by Eq. (\ref{epsilon}).

The approximate effective equations (\ref{syst}) for the $\phi$ field are  similar to the system of equations for the conventional slow roll. This similarity suggests that there should be a redefinition of the field that transforms system (\ref{syst}) into a system described by GR dynamics.
For the simple NMDC such  a correspondence between the non-minimally coupled inflaton  and the minimally coupled inflaton has been found \cite{Dalianis:2014nwa, Dalianis:2016wpu}.  On the same footing, for the GNMDC  we find that there is also a {\itshape generic} transformation of the form
\begin{align} \label{trans1}
\varphi=g(\phi)~, \quad\quad V_m(\varphi)=V[g^{-1}(\varphi)]~,
\end{align}
such that the above system of equations (\ref{syst}) is recast into
\begin{equation} \label{msyst}
H^2\simeq \frac{V_m(\varphi)}{3M^2_P},\quad\quad 3H\dot{\varphi} \simeq -V'_m(\varphi)\,,
\end{equation}
where $V_m(\varphi)$ is a potential for the new field $\varphi$, that is minimally coupled to gravity.
Following these steps the EOM of (\ref{msyst}) is written in terms of the $\phi$ field as $3H\dot{\phi} \simeq -{V'(\phi)}/{[g'(\phi)]^2}$
where $g'(\phi)=d\varphi/ d\phi$.  After straightforward algebra the form of the system (\ref{syst}) is obtained.
For ${\cal A} \gg {\cal B}$ it is $g'(\phi)=(\epsilon_V/\epsilon)^{1/2} \approx {\cal A}$.
Therefore the new field $\varphi$ reads in terms of the field $\phi$,
\begin{equation} \label{trans}
\varphi \,= \, \int \left(\frac{\epsilon_V}{\epsilon}\right)^{1/2} d\phi \, = \,\frac{1}{M_\text{Pl}}  \int   \left[\, {V^{}(\phi)}\, f^{}(\phi) \, \right]^{1/2} \,d\phi\,.
\end{equation}

Summarizing, the dynamics of a field $\phi$ with GNMDC and potential $V(\phi)$ in the slow roll regime of the HF limit, defined by the system (\ref{syst})  are equivalently described, up to first order in the slow roll parameters, by a canonical field $\varphi$, given by Eq. (\ref{trans}), with potential $V_m(\varphi)$,  given by Eq. (\ref{trans1}) and GR gravity.
In the next subsections we analyze  prototype forms for the potential, discussed already in the previous section, focusing on the power law monomial and exponential potentials for simple and motivated forms for the GNMDC,
\begin{align}
f(\phi)\, = \, \alpha \frac{\phi^{\alpha-1}}{M^{\alpha+1}} \,~, \quad \text{or} \quad f(\phi)\, = \, \alpha \frac{e^{2\tau \phi/M_\text{Pl}}}{M^2}~.
\end{align}
These $f(\phi)$ forms are useful for our analytical purposes as well, since it is the product $V(\phi) \cdot f(\phi)$ that appears in transformation (\ref{trans}).

\subsubsection{Correspondence for the power law monomial potentials}
Let us consider here the general class of power law monomial potentials, with a GNMDC of similar form
\begin{equation} \label{monom}
V(\phi)= \lambda_p \,\phi^p\,, \quad\quad f(\phi)\, = \, \alpha \frac{\phi^{\alpha-1}}{M^{\alpha+1}}\,,
\end{equation}
where $\lambda_p$ has dimensions [mass]$^{4-p}$.
According to the transformation (\ref{trans}), $\varphi=\int (\epsilon_V/\epsilon)^{1/2} d\phi$,
where
\begin{align}
 \frac{\epsilon_V}{\epsilon} \, =\, \left( 3 H^2 \, \frac{\alpha\, \phi^{\alpha-1}}{M^{\alpha+1}} \right) \, \simeq  \,  \left( \frac{\alpha V}{M^2_P M^2} \frac{ \phi^{\alpha-1}}{M^{\alpha-1}} \right) \gg 1\,,
\end{align}
and the field $\varphi$  is given by the expression
\begin{equation}
\varphi =  2\,\lambda^{1/2}_p \left(\frac{\alpha}{M^{\alpha+1}} \right)^{1/2} \frac{\phi^{\alpha+p+1}}{(\alpha+p+1)/2}\frac{1}{M_\text{Pl}}~.
\end{equation}
$\varphi$ acts as minimally coupled to gravity during the inflationary phase, governed by
 the standard slow-roll regime system of equations (\ref{msyst}).

The potential $V_m(\varphi)=V[g^{-1}(\varphi)]$ is given by the function,
\begin{equation} \label{vmm}
V_m(\varphi) = \lambda_p \, \left[     \frac{\alpha+p+1}{2}\frac{M_\text{Pl}}{ \lambda^{1/2}_p}  \,  \left(\frac{M^{\alpha+1}}{\alpha}\right)^{1/2}\, \varphi   \,    \right]^{\frac{2p}{\alpha+p+1}}\,,
\end{equation}
where we plugged expession $g^{-1}(\varphi)=\phi$ in Eq. (\ref{monom}).

Thus, we find that during slow-roll there is a direct correspondence between the potential $V(\phi)$ for the non-minimally kinetically coupled inflaton and the $V_m(\varphi)$ for the minimally coupled inflaton
\begin{equation} \label{power-cor}
V\propto \phi^p \quad \longleftrightarrow \quad V_m\propto \varphi^{\frac{2p}{p+\alpha+1}}\,.
\end{equation}
We observe that for any value $\alpha\geq 1$ the effective
 minimal potential $V_m$
 is always less steep with power less than two.   
Hence potentials $V(\phi)$ with power $p\gg1$ appear effectively as potentials with mild slope that can give viable inflation.
Turning to the number of $e$-folds, fields $\phi$ and $\varphi$ implement an equal amount of inflation,
thus in both pictures $N$ is the same, see equation (\ref{Ntwo}).
We also mention that the excursion for the canonical field $\varphi$ is superplanckian.

Let us overview separately the standard examples.
\\
\\
{\it Quadratic potential}. For the prototype inflationary model  $V=m^2 \phi^2/2$ we find the correspondence 
\begin{align}
V(\phi)=\frac{1}{2}m^2 \, \phi^2\quad \longleftrightarrow \quad V_m(\varphi)=V[\phi(\varphi)]=\frac12 m^2 \left( \frac{ \alpha+3}{2m}  M_\text{Pl}  \sqrt{\frac{2M^{\alpha+1}}{\alpha}}  \right)^{\frac{4}{\alpha+3}} \varphi^{\frac{4}{\alpha+3}}~.
\end{align}
Hence, the quadratic potential is mapped into the monomial potential with power $\varphi^{4/(\alpha+3)}$.
\\
\\
{\it Quartic potential}.  Let us look into the quartic Higgs-like potential. We find that
\begin{equation}
V(\phi)= \lambda \, \phi^4\quad \longleftrightarrow \quad V_m (\varphi)
=\lambda\, \left( \frac{\alpha+5}{2} \frac{M_\text{Pl}}{\lambda^{1/2}}\frac{M^{\alpha+1}}{\alpha} \right)^\frac{8}{\alpha+5}\, \varphi^\frac{8}{\alpha+5},
\end{equation}
i.e. the quartic potential for an inflaton with non-minimal kinetic coupling is equivalent to $\varphi^{8/(\alpha+5)}$ monomial potential for an inflaton with minimal coupling.

{\it Linear potential}.  For the linear potential there is the  correspondence,
\begin{align}
V(\phi)= m^3 \, \phi \quad \longleftrightarrow \quad V_m(\varphi)=m^3  \left( \frac{ \alpha+2}{2\sqrt{\alpha}} \frac{M_\text{Pl}}{m^{3/2}} M^\frac{\alpha+1}{2} \right)^{\frac{2}{\alpha+2}} \varphi^{\frac{2}{\alpha+2}}~,
\end{align}
see also ref. \cite{McAllister:2014mpa} for relevant monomial potentials in stringy and \cite{Ferrara:2014fqa} in supergravity set ups.

\subsubsection{Correspondence for the exponential potentials}

Let us consider the exponential potentials $V(\phi)$, with an exponential GNMDC function $f(\phi)$,
\begin{equation} \label{expo}
V(\phi)=V_0 e^{2\lambda \phi/M_\text{Pl}}\,, \quad\quad f(\phi)\, = \,  \frac{e^{2\tau \phi/M_\text{Pl}}}{M^2} \,.
\end{equation}
For ${\cal A} > {\cal B}$, that is for ${\cal A} > \lambda \tau$ the canonical field $\varphi$ reads in terms of the non-minimally coupled field $\phi$
\begin{align}
\phi \simeq -\frac{M_\text{Pl}}{\tau+\lambda} \, \ln \left[(\tau +\lambda)\,\frac{M}{V_0^{1/2}} \, \varphi\right]\,.
\end{align}
The corresponding GR potential has the remarkably simple monomial power law form
\begin{align} \label{expon2}
V_m(\varphi)\, = \, V[\phi(\varphi)] \,= \, V_0 \left[ \frac{(\tau+\lambda)}{V^{1/2}_0} M_\text{Pl} \right]^{\frac{2\lambda}{\lambda+\tau}}\, \varphi^{\frac{2\lambda}{\lambda+\tau}} \,.
\end{align}
Moreover, for $\lambda^2>1/2$, inflation terminates automatically when the GNMDC becomes ineffective, contrary to the GR case, where inflation with an invariant exponential potential is endless \cite{Lucchin:1984yf, Yokoyama:1987an}.
The graceful exit is realized thanks to the fact that  ${\cal A}\rightarrow 1$ as both  the Hubble scale $H$ and the field value $\phi$ decrease.
Afterwards, a (quasi) kination stage, with $w\simeq 1$, is realized.

Summarizing,  the exponential potentials with a GNMDC of exponential form give similar inflationary observables with monomial potentials $\varphi^{n}$ with $n<2$, in accordance with the Planck 2018 constraints! Namely,
\begin{align}
V(\phi)\, =\, V_0 \, e^{2\lambda \phi/M_\text{Pl}} \quad \longleftrightarrow \quad V_{m}(\varphi) \propto  \varphi^{\frac{2\lambda}{\lambda+\tau}}\,.
\end{align}
and for $\tau>1$ the exponential potentials  are actually  very attractive inflationary models in the context of GNMDC.
We can say that GNMDC revives the inflationary exponential potentials, that additionally feature a graceful exit from inflation as well.

Let us finally briefly comment on the
inflationary phenomenology
when the potential is
 $V(\phi)=V_0 e^{2\lambda \phi/M_\text{Pl}}$ and the coupling to $G_{\mu\nu}$ is given by the function $f(\phi)=\alpha \phi^{\alpha-1}/M^{\alpha+1}$.
 The correspondence between the non-canonical $\phi$ and the canonical $\varphi$ fields is found to be
\begin{align}
\varphi=- 2^{\frac{1+\alpha}{2}} \, \sqrt{a}\, \frac{{V_0}^{1/2}}{M_\text{Pl}}
 \left(-\frac{M_\text{Pl}}{2\lambda \, M} \right)^{\frac{1+\alpha}{2}}  \Gamma\left[\frac{1+\alpha}{2}, -\frac{2\lambda \phi}{M_{P}}\right]~,
\end{align}
where $\Gamma$ is the incomplete Gamma function, and for $\phi>0$  odd values for $\alpha$ are required. The analytic expression for the $V_{m}(\varphi)$ function can be found only for $\alpha=1$ and reads $V_{m}(\varphi)=\left( {\lambda M}\right)^{2} \varphi^2$. For $\alpha>1$ only approximate expression can be found and only for odd values for $\alpha$.

\begin{figure}[htbp]
\makebox[\textwidth]{
\includegraphics[width=0.45\textwidth]{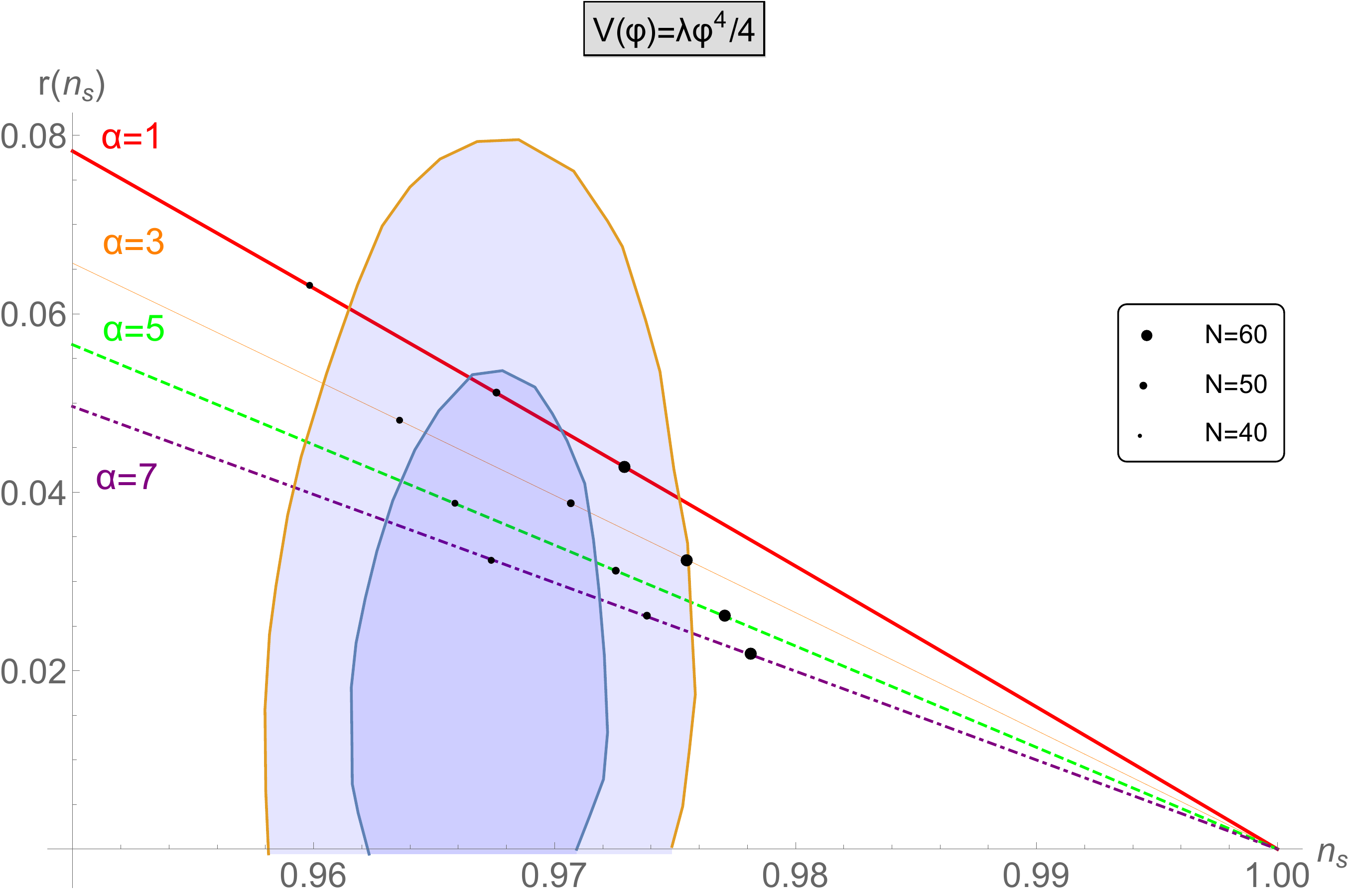}%
\hfill
\includegraphics[width=0.45\textwidth]{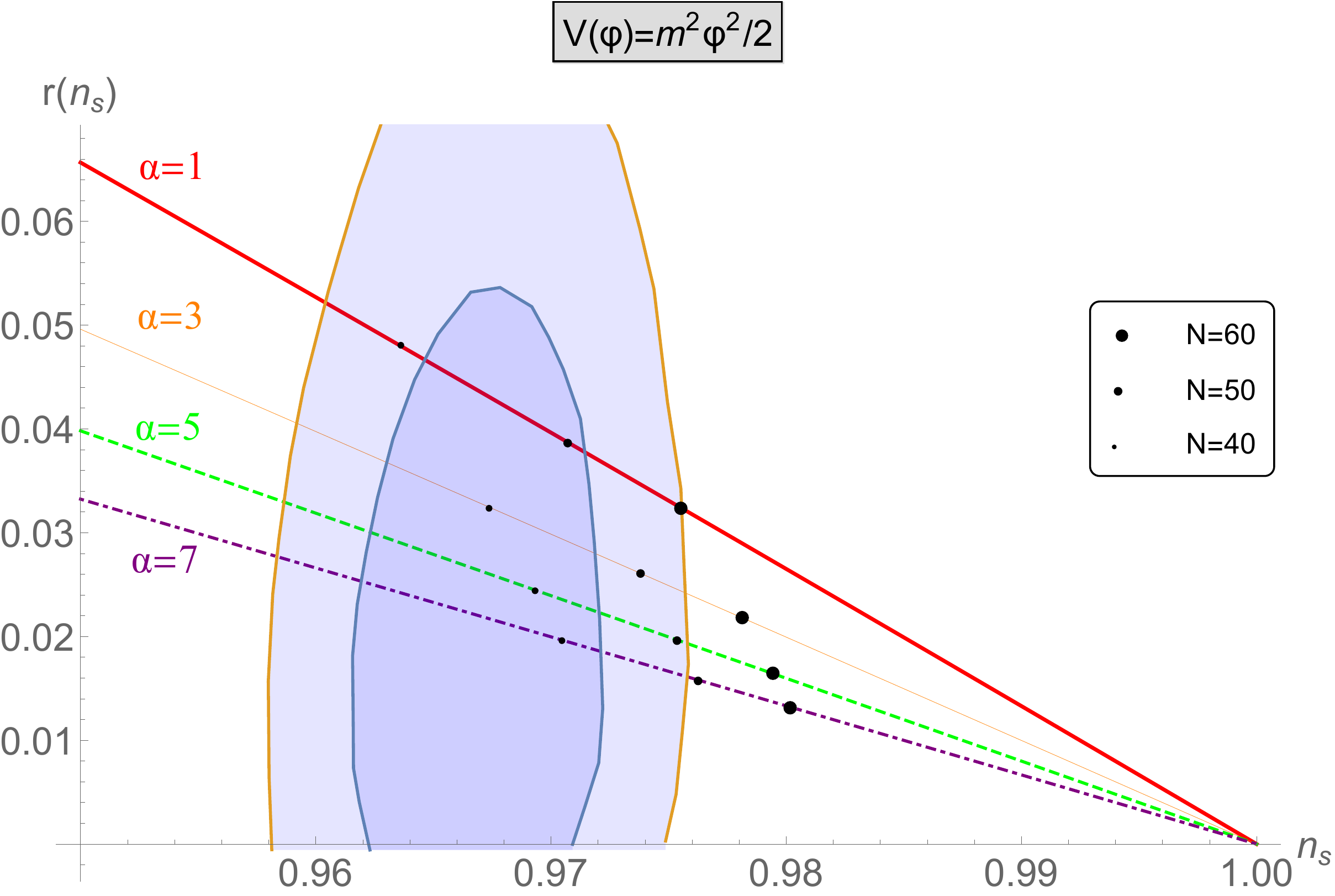}%
}\\[0.4cm]

\caption{The $r(n_s)$ contour plot for a Higgs ({\it left panel}) and a quadratic ({\it right panel}) potential, for $\alpha=1$ (red), $\alpha=3$ (orange), $\alpha=5$ (green), $\alpha=7$ (purple) against the 68\% and 95\% CL regions of Planck 2018 \cite{Akrami:2018odb}. The bullet points correspond, as shown in the legends, to particular numbers of e-folds obtained by numerical methods. 
}
\label{fig:rofnsplots}
\end{figure}

\begin{figure}[htbp]
\makebox[\textwidth]{
\includegraphics[width=0.45\textwidth]{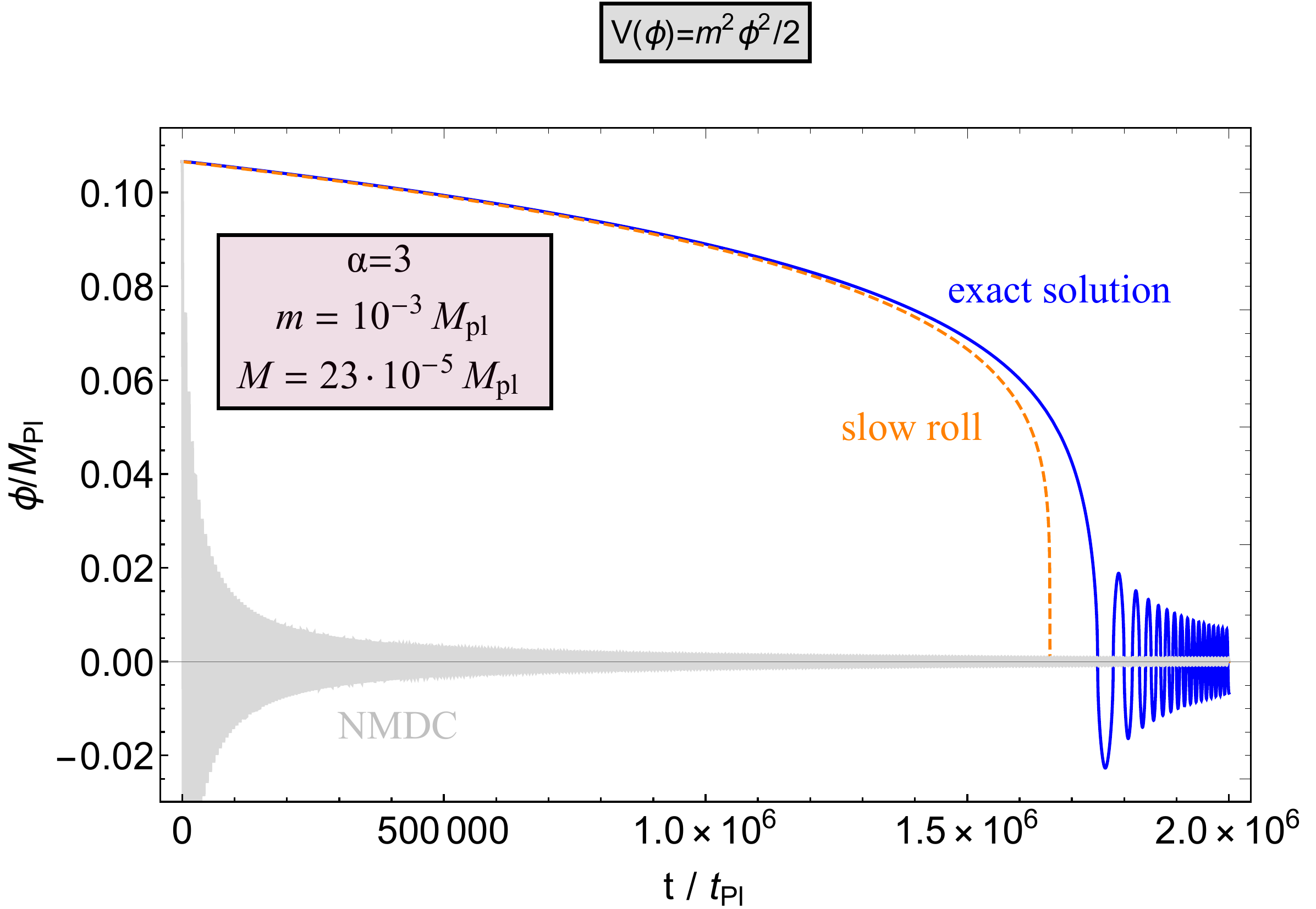}%
\hfill
\includegraphics[width=0.45\textwidth]{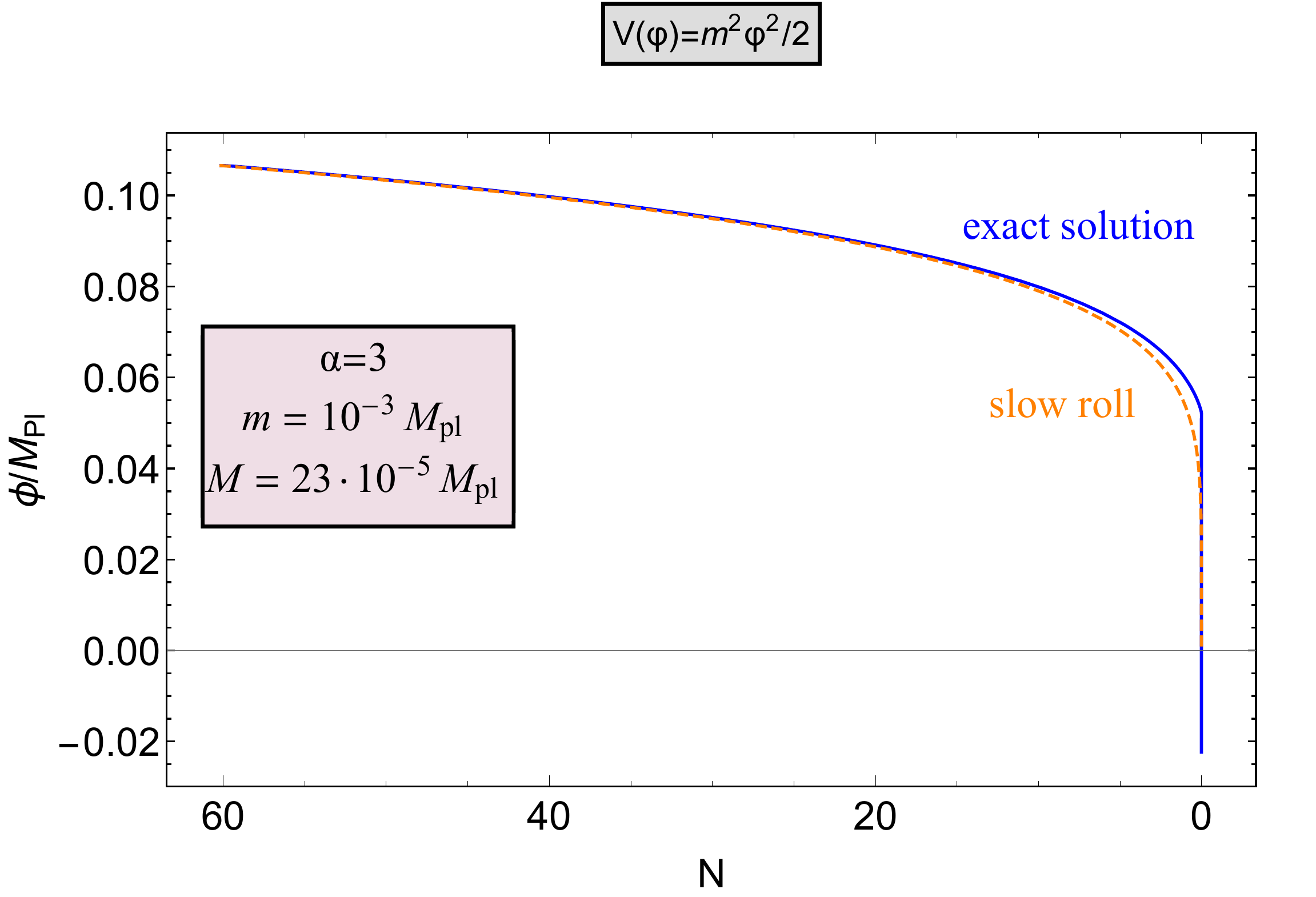}%
}\\[0.4cm]

\caption{.
The evolution of the scalar field  $\phi(t)$ ({\it left panel}) and $\phi(N)$ ({\it right panel}) for $\alpha=3$  and a quadratic potential. In the plots both the exact and the slow-roll approximation solutions are depicted. Again, the slow-roll approximation proves particularly accurate early on. The parameters of this model have been chosen so that it yields observables in accordance with Planck 2018. 
For simple reference, the evolution of the model with the same  parameters and  $\alpha=1$ (NMDC), is also included.
}
\label{fig:examplequad}
\end{figure}

\begin{figure}[htbp]
\makebox[\textwidth]{
\includegraphics[width=0.45\textwidth]{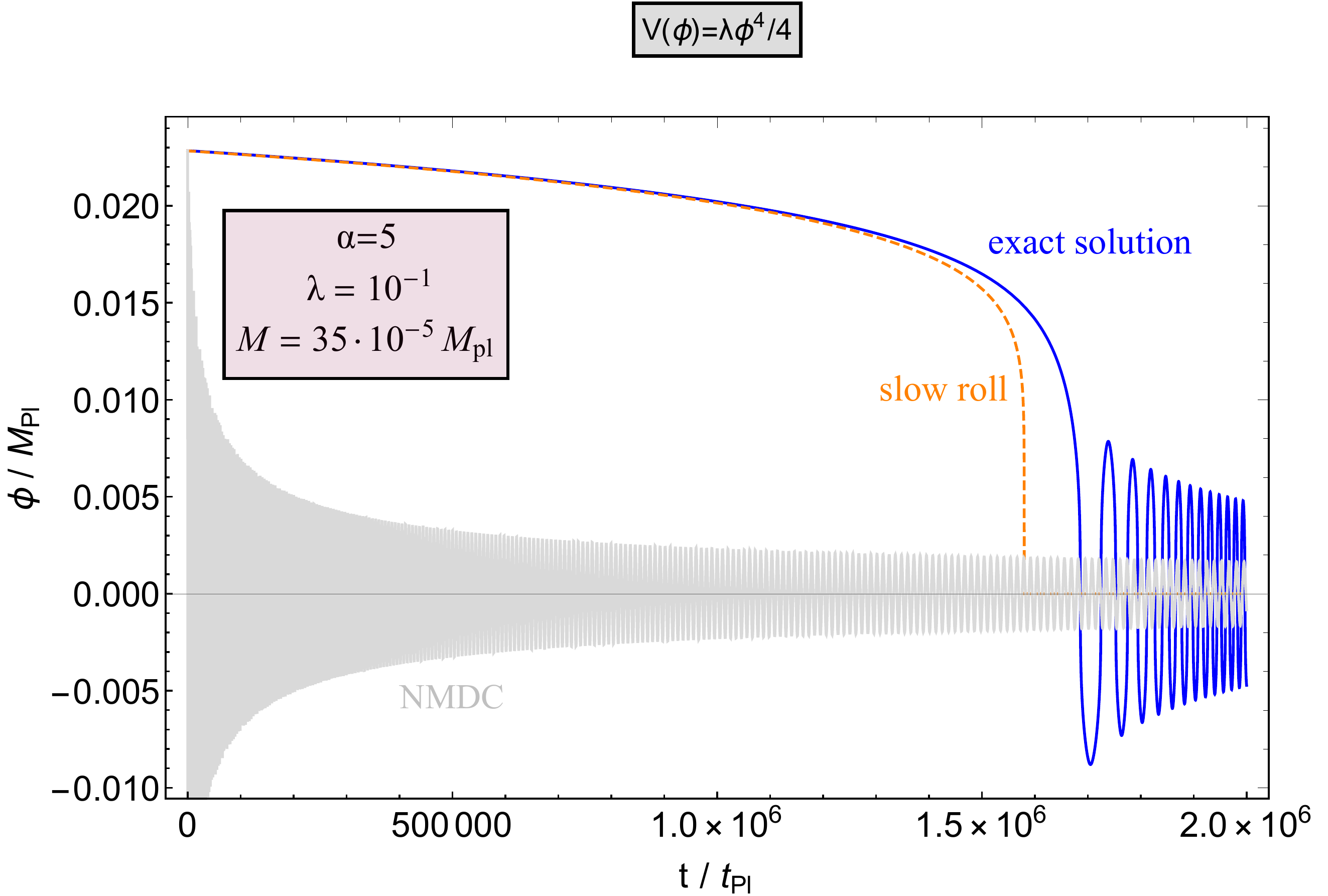}%
\hfill
\includegraphics[width=0.45\textwidth]{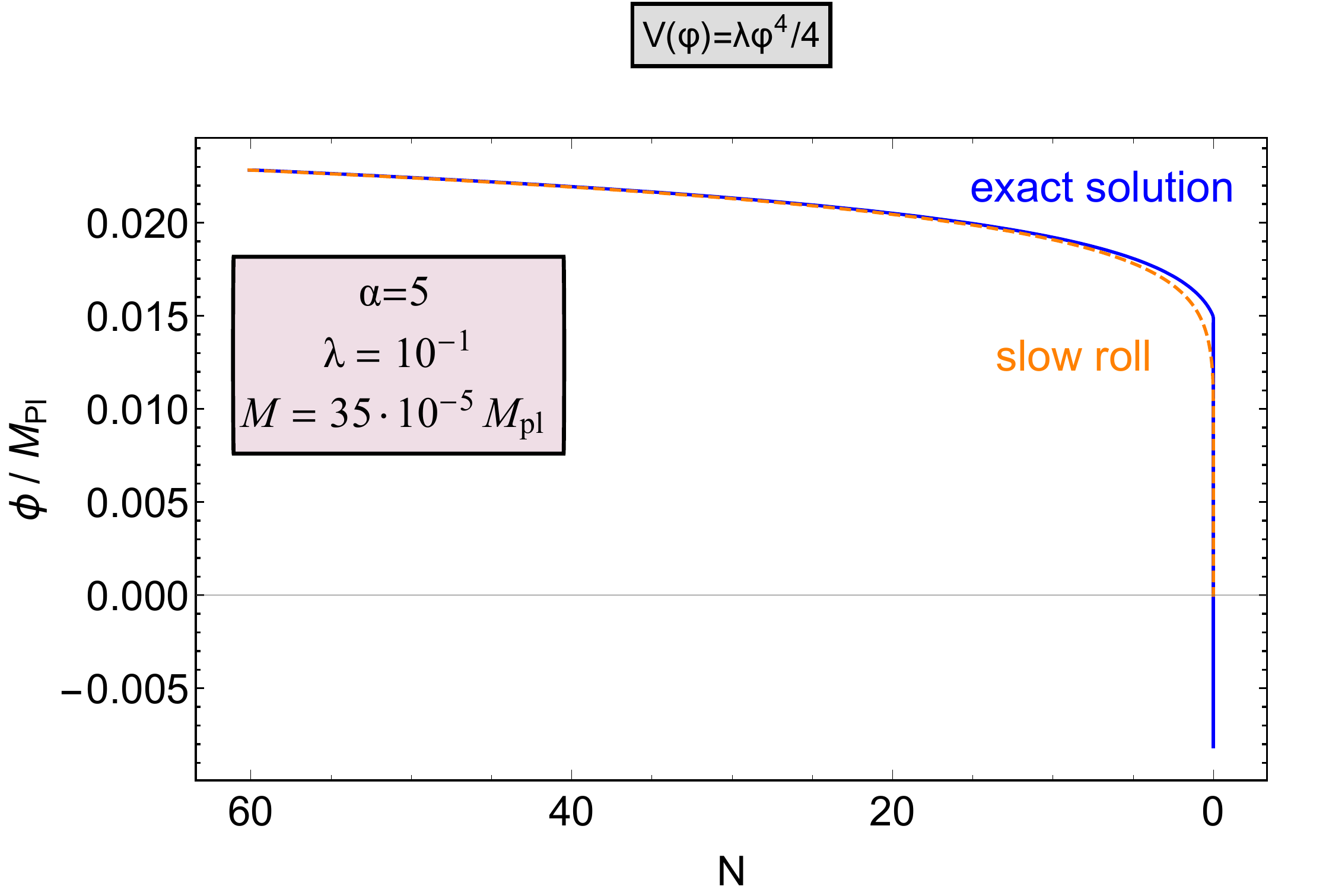}%
}\\[0.4cm]

\caption{As in Fig. \ref{fig:examplequad},  the evolution of the $\phi(t)$ ({\it left panel}) and $\phi(N)$ ({\it right panel}) for $\alpha=5$ for the  Higgs potential is depicted. 
}

\label{fig:examplehiggs}
\end{figure}

\subsection{The expansion history after inflation with GNMDC}

The analysis of the previous subsection revealed that there is a phenomenological  correspondence between GNMDC and GR.
For example,  for  the power-law potentials the correspondence  reads (see \eqref{power-cor})
\begin{equation} \label{corre2}
\left.  p \,\Bigr|_\text{GNMDC} \,\longleftrightarrow \,\frac{2p}{p+\alpha+1}\right|_\text{GR}\,\equiv q\,.
\end{equation}
Utilizing this correspondence, the basic inflationary predictions, derived by  examining solely the GNMDC dynamics, given by Eq. (\ref{ns-r}) and (\ref{nsexp}) can be reproduced after utilizing the
well-known GR relations $1-n_s\simeq (2q+4)/(4N+q)$ and $r\simeq 16 q/(4N+q)$.
For this, we assumed that the same number of $e$-folds take place both in the corresponding GR mode and in the original GNMDC model.
Indeed, in the HF limit the number of $e$-folds are approximately equal
\begin{align} \label{Ntwo}
N \, \simeq 
 \, \frac{1}{M^2_\text{Pl}} \, \int_{\phi_\text{end}}^{\phi} [g'(\phi)]^2 \, \frac{V}{V'} \, d\phi \,
 = \, \frac{1}{M^2_\text{Pl}} \, \int_{\varphi_\text{end}}^{\varphi} \frac{V_m}{V'_m}  \, d\varphi \,,
\end{align}
where we took into account that $g'(\phi)=d\varphi/ d\phi$ and $(dV/d\phi) (d\phi/ d\varphi)=dV_m/d\varphi$.

The number of $e$-folds are critical for the inflationary model selection, see Eq. (\ref{ns-r}) and (\ref{nsexp}). The exact number of $e$-folds  $N$ depend on the postinflationary evolution of the universe that is highly unknown due to the absence of any cosmic observables preceding the BBN.
Nevertheless, each inflationary model admits a restricted range of $N$-values   unless complicated cosmological scenarios are adopted.
After the end of the acceleration phase, the inflaton is expected to oscillate about the minimum of its potential,
gradually transforming its energy to other degrees of freedom.
For power-law potentials $V(\phi)\propto \phi^p$, the averaged effective equation of state for a coherently oscillating scalar field with GR dynamics  is given by the well-known relation   \cite{Turner:1983,Shtanov:1994ce},
\begin{equation} \label{potw1}
{w}_\text{} \equiv \frac{\left\langle p\right\rangle}{\left\langle \rho\right\rangle}=\frac{p-2}{p+2}\,.
\end{equation}
The reheating period leads to the thermalization of the universe
and the initiation of the radiation era with $T_\text{reh}\sim ({\Gamma_\phi M_\text{Pl}})^{1/2}$, where $\Gamma_\phi$ is the inflaton decay rate.
The number of $e$-folds of observable inflation are
\begin{equation} \label{Nrh}
N_\text{} \simeq 57.6 +\frac14 \ln \epsilon_* +\frac14 \ln \frac{V_*}{\rho_\text{end}} -\frac{1-3w_\text{}}{4} \tilde{N}_\text{rh} \,,
\end{equation}
where  $\tilde{N}_\text{rh}$ the $e$-folds that take place during reheating.
Thus, $N$ has also a dependence on the reheating stage through the parameters $\tilde{N}_\text{rh}$ and $w$. 

GNMDC models have a very distinct reheating stage.
 There are two basic cases: i) GNMDC dominates over GR dynamics during the reheating stage and, ii) GR dynamics take over the GNMDC dynamics after the end of inflation.
In the former case, GNMDC strongly modifies the relation (\ref{potw1}) \cite{Dalianis:2016wpu} predicting distinct values for the  factor $(1-3w)\tilde{N}_\text{rh}/4$.  In the latter case, GNMDC becomes ineffective during the reheating stage and Eq. (\ref{potw1}) applies. Also here there is a clear distinction with the GR models. While the reheating equation of state is determined by the potential $V(\phi)$, the inflationary dynamics are determined by both the potential $V(\phi)$ and the coupling $f(\phi)$ to $G_{\mu\nu}$.
To be specific, let us briefly overview the reheating predictions for the  Higgs and  exponential inflationary models discussed in this paper.

\begin{itemize}
\item The Higgs inflation with potential $V(\phi)=\lambda \phi^4/4$ and GNMDC function $f(\phi)\propto  \phi^{\alpha-1} $ predicts $n_s$ and $r$ values given by eq. (\ref{ns-r}), after substituting $p=4$. Let us assume a benchmark value  $\alpha=11$. The Planck 2018 spectral index and tensor-to-scalar ratio values are   found to be $n_s=0.965$ and $r=0.054$ for $N=36$. The $36$ $e$-folds
imply that  an extended non-thermal phase takes place before the BBN epoch. The duration of this non-thermal phase can be specified.
After a few oscillations the GR dynamics dominate and the equation of state  after inflation is $w=1/3$ regardless of the exact reheating temperature value, as Eq. (\ref{potw1}) dictates. 
 On the other hand, the GR model that predicts the same $n_s$ and $r$ values for the same number of $e$-folds has a potential   $V(\varphi)\propto \varphi^{1/2}$. Hence, according to Eq. (\ref{potw1}) the postinflationary equation of state is different and in principle the two models are distinguishable.  A similar argumentation applies also for any power-law monomial potential with GNMDC.

\item Inflation with exponential potential, $V(\phi)=e^{2\lambda \phi/M_\text{Pl}}$, and GNMDC function $f(\phi)\propto  e^{2\tau \phi/M_\text{Pl}}$ is another interesting example.
 Let us assume benchmark values $\lambda=100$,    $\tau=320$.  For these numbers the relations (\ref{nsexp}) apply since it is ${\cal A} \gg {\cal B}$.
We find that $n_s=0.969$ and $r=0.048$ for $N=40$ $e$-folds.
The number of $e$-folds implies 
the existence of a non-thermal postinflationary phase additional to the kination  era that follows inflation with exponential potential.
The corresponding GR model that predicts the same $n_s$ and $r$ values for the same number of $e$-folds has a potential   $V(\varphi)\propto \varphi^{10/21}$ and therefore eq. (\ref{potw1}) indicates  a different  postinflationary evolution  and  the two models are, again, distinguishable.
\end{itemize}

A thorough analysis of the inflationary and reheating dynamics with NMDC can be found in Ref. \cite{Dalianis:2016wpu}.

\section{PBH production from the GNMDC} \label{PBHsec}
\label{SecPBH}

\subsection{Preliminaries on PBHs}\label{PrePBH}

In this section we turn to PBHs cosmology; for a review and related works see Ref. \cite{Sasaki:2018dmp} and \cite{Khlopov:2008myu, Raidal:2018bbj, Carr:2017jsz} respectively. Similar attempts have also been made in $f(R)$ scenarios \cite{Pi:2017gih} and trapped inflation \cite{Cheng:2016qzb}.
We will specify  the required magnitude of the power spectrum amplification and the inflationary stage during which this  amplification occurs.
Roughly speaking, the former determines the  abundance of the PBHs  and the latter the mass of the PBHs produced.

\subsubsection{The amplitude of the power spectrum peak}
PBHs with mass $M$ form
 due to large density perturbations that  gravitationally dominate over the radiation pressure and collapse after the horizon reentry.
The PBH mass is equal to $\gamma M_\text{hor}$ where $M_\text{hor}$ is the horizon mass and $\gamma$  a numerical factor which depends on the details of the gravitational collapse.
The present ratio of the abundance of PBHs  with mass $M$ over the total dark matter (DM) abundance,  $f_\text{PBH}(M)\equiv {\Omega_{\text{PBH}}(M)}/{\Omega_{\text{DM}}}$,  is expressed as
 \begin{equation} \label{fpbh}
f_\text{PBH} (M)\, =\,\left(\frac{\beta_\text{}(M)}{7.3 \times 10^{-15}}\right) \,
\left(   \frac{\Omega_{\text{DM}}h^2}{0.12}   \right)^{-1}
 \Big(\frac{\gamma_\text{}}{0.2}\Big)^{\frac{3}{2}}
\left(\frac{g(T_{})}{106.75}\right)^{-\frac{1}{4}}
\left(\frac{M}{10^{20}\text{g}}\right)^{-1/2}\,,
\end{equation}
where we took the effective degrees of freedom $g_*$ and $g_s$ approximately equal.
 The theory for the PBH formation that we follow is based on the traditional Press-Schechter formalism \cite{Press}.  
This consideration is regarded as the conventional one for PBH formation,
for a different recent suggestion  see Ref. \cite{Germani:2018jgr}. 
 We assume that curvature perturbations are described by   Gaussian statistics in order to estimate the PBH formation probability and  connect the collapse threshold to the power spectrum.  For spherically symmetric regions PBHs form with rate $\beta$,

\begin{equation} \label{brad}
\beta_\text{}(M)=\int_{\delta_c}d\delta\frac{1}{\sqrt{2\pi\sigma^2(M)}}e^{-\frac{\delta^2}{2\sigma^2(M)}}\,
\simeq \, \frac{1}{2}\text{erfc}\left(\frac{\delta_c}{\sqrt{2}\sigma(M)}\right)
\, \simeq \,  \frac{1}{\sqrt{2\pi}} \frac{\sigma(M)}{\delta_c} e^{-\frac{\delta^2_c}{2\sigma^2(M)}}  \,.
\end{equation}
Parameter $\delta_c$ is the threshold density perturbation and erfc$(x)$ is the complementary error function.  For $\delta>\delta_c$, density perturbations overcome internal pressure and collapse.

The PBH formation rate depends on the variance of the density perturbations.
The variance of the density perturbations $\sigma(k)$ smoothed on a scale $k$ for radiation domination is given by \cite{Young:2014ana}
\begin{equation}
\sigma^2(k)= \left( \frac{4}{9} \right)^2  \int \frac{dq}{q}W^2(qk^{-1})(qk^{-1})^4{\cal P_R}(q)\,,
\end{equation}
where ${\cal P_R}(q)$ is the power spectrum of the curvature perturbations. The $W(z)$ represents the Fourier transformed function of the Gaussian window,  $W(z)=e^{-z^2/2}$. An order of magnitude estimation of the $\beta(M)$ can be done after the approximation  ${\cal P_R}(k) \simeq (9/4)^2\sigma^2(k)$,
\begin{align} \label{bP}
\beta(M)_\text{}\,  \sim \, \frac{1}{\sqrt{2\pi}} \frac{\sqrt{{\cal P}_{\cal R}}}{\delta_c} \,\,
e^{-{\delta^2_c}/{2{\cal P}_{\cal R}} }   \,.
\end{align}

The power spectrum in the GNMDC gravity models might be very sensitive to the $\phi$ field value.  According to Eq. (\ref{PR}) it is,
 \begin{align} \label{PRpbh}
{\cal P}_{{\cal R}}	\approx \frac{V}{96\pi^2 M_\text{Pl}^4 \epsilon_V/{\cal A}}~.
\ 
\end{align}
Apparently, when $\epsilon$ decreases, $\beta(M)_\text{}$ increases.
Asking for $f_\text{PBH}\sim 1$, the power spectrum has to be  ${\cal P}_{\cal R}^\text{PBH} \sim 10^{-2}$ for $\delta_c \sim 0.5$.
From (\ref{PRpbh}) it is $\epsilon(\phi_\text{PBH})=V(\phi_\text{PBH})(96 \pi^2 M^4_\text{Pl})^{-1} {\cal (P_R^\text{PBH}})^{-1}$.
Since $V(\phi_\text{PBH})<V_\text{max}=3\pi^2 A_{s} r_\text{max}\, M^4_\text{Pl}/2$ the $\epsilon_\text{PBH}$ has to be smaller than a value, $\epsilon_\text{max}$, that is given when we substitute $A_s\simeq 2.18 \times 10^{-9}$ and $r_\text{max}\simeq 0.64$,  from Planck 2018 data  \cite{Akrami:2018odb}, and reads,
\begin{align}
\epsilon_\text{PBH}\, \lesssim
 \frac{10^{-11}}{{\cal P}_{\cal R}^\text{PBH}} \sim    \, 10^{-9}\,.
\end{align}
The dramatic decrease of $\epsilon$ implies that $\delta$ and $\eta$ increase and the slow-roll approximation is violated, hence one has to solve the Mukhanov-Sasaki equation, see next sections.

\subsubsection{The PBH mass}

For a scale $k^{-1}_\text{}$,  which exits the Hubble horizon after
 $N_{k\text{}}$  e-folds  before the end of inflation,
there is the relation
\begin{equation} \label{Dnk}
 N_k= \ln\left(\frac{k_\text{end}}{k}\right)-\ln\left(\frac{H_\text{end}}{H_k}\right)\,.
\end{equation}
Let us assume that  $H_\text{end}\simeq H_k $ which is a very good approximation for scales $k^{-1}$ that exit  the Hubble horizon during or after the ultra slow-roll inflationary phase \cite{Kinney:2005vj,Motohashi:2017kbs}, so the second term in Eq. (\ref{Dnk}) can be neglected.
After the end of inflation the Hubble horizon, $H^{-1}$, grows fast 
and the scales gradually reenter the horizon.
We define $\tilde{N}_{k}\equiv \ln (a(t)/a_\text{end})$ the e-folds that take place after the end of inflation until reentry, denoted with a tilde to make a clear distinction  with the e-folds that take place during inflation.  For $w>-1/3$, it is
$\tilde{N}_{k\text{}}=2\,N_{k_\text{}}/{(1+3w)}\, $.
Unless $w = 1/3$, the number of $e$-folds at which a particular scale reenters the horizon depends on the reheating temperature.
If a perturbation with scale $k^{-1}$ enters inside the horizon during a phase with equation of state $w=1/3$, that is during radiation domination, or during $\phi^4$ oscillating stage, the relation between the scale $k^{-1}$ and the horizon mass $M/\gamma$ reads
\begin{align} \label{kRDm}
k_\text{}(M)\, = \, 1.8 \times 10^{18}\, \text{Mpc}^{-1}\,  \gamma^{1/2}_\text{} \,  \left(\frac{M}{10^{10}\, \text{g}} \right)^{-1/2}   \left(\frac{g_*}{106.75} \right)^{-1/12} \,.
\end{align}
Let us focus on two especially interesting mass scales:
\begin{enumerate}

\item[(i)] $M_\text{PBH} \simeq 10^{21}$ g : PBHs in this mass range  can explain all of the  dark matter density observed in the universe today. Interestingly enough, apart from direct observational tests,  scenarios with PBHs of that mass can be probed by gravitational wave experiments such as LISA \cite{Garcia-Bellido:2017aan, Cai:2018dig, Bartolo:2018rku, Cai:2019jah}.

\item [(ii)] $M_\text{PBH}\simeq 10^{35}$ g : PBHs of that mass can explain the BH events observed by LIGO \cite{Abbott:2016blz}. In this mass range PBHs cannot constitute all of the dark matter, nevertheless they can comprise a significant fraction, of the order of few percent \cite{Carr:2016drx}.
\end{enumerate}

If PBHs with mass $M_\text{PBH}=10^{21}$ g enter the horizon during a cosmic phase with equation of state $w=1/3$ it is $k(10^{21}\, \text{g})=5.7 \times 10^{12} \, \text{Mpc}^{-1} \gamma^{1/2}=10^{14} \, k_\text{CMB}\, \gamma^{1/2}$.
For   $M_\text{PBH}=10^{35}$ g we get  $k(10^{35}\, \text{g})=5.7 \times 10^{5} \, \text{Mpc}^{-1} \gamma^{1/2} =1.1 \times 10^{7} \, k_\text{CMB}\, \gamma^{1/2}$. We took $g_* \simeq {\cal O}(100)$ and we explicitly labeled $k_\text{CMB}$ the CMB pivot scale, $k_\text{CMB}=0.05$ Mpc$^{-1}$ to make the distinction with the scale $k_\text{PBH}$ clear.

The next step is to determine the $e$-folds $ N_k=N_\text{PBH}$ before the end of inflation, at which the peak in the power spectrum is generated.
$N_\text{CMB}$ is related to the
postinflationary cosmic expansion rate via expression (\ref{Nrh}).
In our models it is $\ln (\epsilon_* V_*/\rho_\text{end})^{1/4} = \pm{\cal O}(1)$.
Hence the above expression constrains $N_\text{CMB} \lesssim 58$ e-folds for $w \leq 1/3$. From relation (\ref{Dnk}) we find $k_\text{end}$,
\begin{align}
k_\text{end}= k_\text{CMB} \, e^{N_\text{CMB}} \, \frac{H_\text{end}}{H_\text{CMB}}~.
\end{align}
For Higgs inflation it is $N_\text{CMB} \simeq 58$ (unless there is an early non-thermal stage) and for our models it is $H_\text{CMB}/H_\text{end}\simeq {\cal O}(2.6)$ thus $k_\text{end}\simeq  4 \times 10^{25} \, k_\text{CMB}$.
Hence, from the relation $N_\text{PBH} \simeq \ln \left( {k_\text{end}}/{k_\text{PBH}}\right)$
 we find that PBHs with mass in the scales (i) and (ii) are produced due to a peak in the power spectrum generated respectively
\begin{align} \label{N2pbh}
N_\text{PBH} (M_\text{PBH}=10^{21} \, \text{g})\, \sim \, 27~, \quad\quad \text{and} \quad\quad N_\text{PBH} (M_\text{PBH}=10^{35} \, \text{g})\, \sim  \,45~,
\end{align}
$e$-folds before the end of inflation.
From Eq. (\ref{fpbh}) we get $f_\text{PBH}(M_\text{PBH}=10^{21} \, \text{g})\sim1$ for $\beta(M)\sim 10^{-13}$ and $f_\text{PBH}(M_\text{PBH}=10^{35} \, \text{g})\sim 0.1$ for  $\beta(M)\sim  10^{-7}$. The formation probability values, $\beta(M)$,  give us an order of magnitude  estimation for the required $\epsilon_\text{PBH}$.

In the following we will construct a friction term that increases strongly at scales much smaller than the CMB scale and makes possible the generation of PBHs.
We also take care not to spoil the CMB inflationary observables.
We will consider the Higgs potential $V(\phi)=(\lambda/4) \phi^4$, with $\lambda \simeq 0.1$, motivated by the very fact that it is realized in nature.

\subsection{Power spectrum amplification in the  GNMDC theories}

The power spectrum in GNMDC gravity,  (Eq. \eqref{PR}), gets amplified when the  parameter $\epsilon_V/{\cal A}$ decreases.
The amplification of the power spectrum induces large density perturbations that might collapse generating PBHs.
The observational constraints on the power spectrum, imply that PBHs are produced in accordance with the observations
 if the ${\cal P_R}(k)$ features a sharp peak, see \cite{Dalianis:2018ymb}. In the GNMDC models this can be achieved if the coupling function $f(\phi)$ gets enhanced about a field value.
For illustration and  model building purposes, we will parameterize the enhancement of $f(\phi)$ by splitting it into two parts,
\begin{align}
f(\phi)=f_I(\phi)\left( 1 + f_{II}(\phi)\right)~. 
\end{align}
$f_I(\phi)$ is the GNMDC function that acts in the beginning of the inflation while
the coupling $f_I(\phi)f_{II}(\phi)$ acts in the middle or towards the end of the inflationary stage.
 Hence, $f_{II}(\phi)$ is a function that peaks at a particular value $\phi=\phi_0$ and nearly vanishes for field values away from $\phi_0$.
Inflation is partitioned between stage I, where $f_{II}(\phi)<1$ and $N_{I}$ $e$-folds take place, and stage II where $f_{II}(\phi)>1$ and $N_{II}$ $e$-folds take place.
There is a freedom in choosing the $f(\phi)$ function and we choose a form
\begin{align}\label{fIIr}
	f_{II}(\phi)=\frac{d}{\sqrt{\left(\frac{\phi-\phi_0}{s M_\text{Pl}}\right)^2+1}}
\end{align}
that has been used in Ref. \cite{Fu:2019ttf}.
Parameters $d, s$ and $\phi_0$ are specified by the requirement that the PBHs generated, have a significant cosmic abundance. In particular we will estimate $d, s$ and $\phi_0$ as a function of the ${\cal P_R}$  amplitude and the $N_\text{PBH}$.
It is
\begin{align}
\left. {\cal P_R} \propto \frac{{\cal A}(\phi)}{\epsilon_V} \simeq \frac{3H^2 f_I(\phi)f_{II}(\phi)}{\epsilon_V} \,  \right|_{\phi=\phi_\text{PBH}}~,
\end{align}
where we defined $\phi=\phi_\text{PBH}$ the field value at which ${\cal P_R}$ maximizes, where it is $f_I(\phi)f_{II}(\phi) \gg f_I(\phi)$. Without loss in precision we can take $\phi_0=\phi_\text{PBH}$.
In the second stage of inflation, where $f_{II}(\phi) > 1$, $\dot{\phi}$ decreases significantly. 
The exact $N_{II}$ value is specified  according to the observational constraints on the PBH abundance. For the field segment that $f_{II}(\phi) > 1$ 
the number of $e$-folds read
\begin{align} \label{NII}
 N_{II} =  \int^{\phi_0+\Delta\phi_{II}}_{\phi_0-\Delta\phi_{II}} \frac{H}{\dot{\phi}} d\phi
\simeq
\frac{1}{M_\text{Pl}}\int^{\phi_0+\Delta\phi_{II}}_{\phi_0-\Delta\phi_{II}} \frac{{\cal A}+{\cal B}/2}{\sqrt{2\epsilon_V}} \, d\phi
\end{align}
We  considered that the second stage of inflation starts at $\phi_0 + \Delta \phi_{II}$ and ends at $\phi_0 - \Delta \phi_{II}$, which is actually a good enough approximation for a sharp $f_{II}(\phi)$.

The conventional approximate result  for the power spectrum (\ref{PS-SR}) is valid only for a softly changing $\epsilon$ \cite{Chongchitnan:2006wx}, or equivalently  $f(\phi)$.
If this is not the case the numeric solution of the exact Mukhanov-Sasaki equation has to be pursued.

\subsection{The Mukhanov-Sasaki equation}

Let us rewrite here the  quadratic action  for the curvature perturbation ${\cal R}$ in the comoving 
gauge, 
\begin{equation}
S_{(2)}\, = \, \frac{M^2_\text{Pl}}{2} \, \int dx^4 a^3 Q_s \left[ \dot{{\cal R}}^2 -\frac{c^2_s}{a^2} (\partial_i {\cal R})^2\right]~,
\end{equation}
where  $c^2_s$ is the sound speed squared and $Q_s$ defined respectively in Eq. (\ref{cs2}) and (\ref{S2}). 
Following \cite{Kobayashi:2019} a new coordinate can be introduced, $dy=(c_s/a) dt=c_s d \eta$.  
Along with the redefinitions 
\begin{align}
u=z{\cal R}, \quad \text{with} \quad z=\sqrt{2}a (c_s Q_s)^{1/2}\equiv a \sqrt{2\tilde{\epsilon}}
\end{align}
a transformed action is obtained 
  that yields the familiar form of the Mukhanov-Sasaki equation\footnote{For an equivalent form of the MS equation, written in terms of the slow roll parameters see \cite{Saito:2008}.}  for the Fourier modes $u_k$,
\begin{align} \label{MS}
	u_k''+\left(c_s^2 k^2-\frac{z''}{z}\right)u_k=0,
\end{align}
where a prime denotes differentiation with respect to the conformal time, $\eta$.
The exact power spectrum is obtained after solving the Mukhanov-Sasaki equation
and computing the $u_k$ values at super-Hubble scales and well after it exits the horizon  and its value freezes out,
\begin{equation} \label{PSMS}
 \left. {\cal P_ R} =\frac{k^3}{2\pi^2}\frac{|u_k|^2}{z^2} \right|_{k\ll aH } \,. 
\end{equation}
The initial conditions for the modes $u_k$ are set by the Bunch-Davies vacuum.
Deep inside the Hubble horizon, $k \gg aH$, the evolution of the $z''/z$ is unimportant because $c^2_s k^2 \gg z''/z$. There, 
 all modes have time independent 
frequencies and Eq. (\ref{MS}) reads $u''_k+c^2_sk^2u_k=0$, giving the Minkowski initial condition $u_k =e^{-ik\tau}/{\sqrt{2k}} $ for the Mukhanov-Sasaki equation. 
Here, we solve separately for the real and imaginary parts of each mode $u_k$.  The Bunch-Davies initial conditions  for the differential equation (\ref{MS}) are
\begin{equation}\label{InitCond}
\text{Re}\left[u_k \right]=\frac{1}{\sqrt{2k}}\,, \quad \text{Im}[u_k]=0, \quad \text{Re}\left[\frac{du_k}{dt}\right]=0\,, \quad \text{Im}\left[\frac{du_k}{dt}\right]=\frac{\sqrt{k}}{a(t_i)\sqrt{2}}\,.
\end{equation}
The $a(t_i)$ is the scale factor value when the mode is deep inside the horizon, $a(t_i)H(t_i)\ll k$.
Eq. \eqref{InitCond}  are the initial conditions we use in order to find the evolution of the $u_k$ mode, several efolds after the horizon exit where the $|u_k/z|$ converges to a constant value.
\begin{figure}[htbp]
\makebox[\textwidth]{
\includegraphics[width=0.44\textwidth]{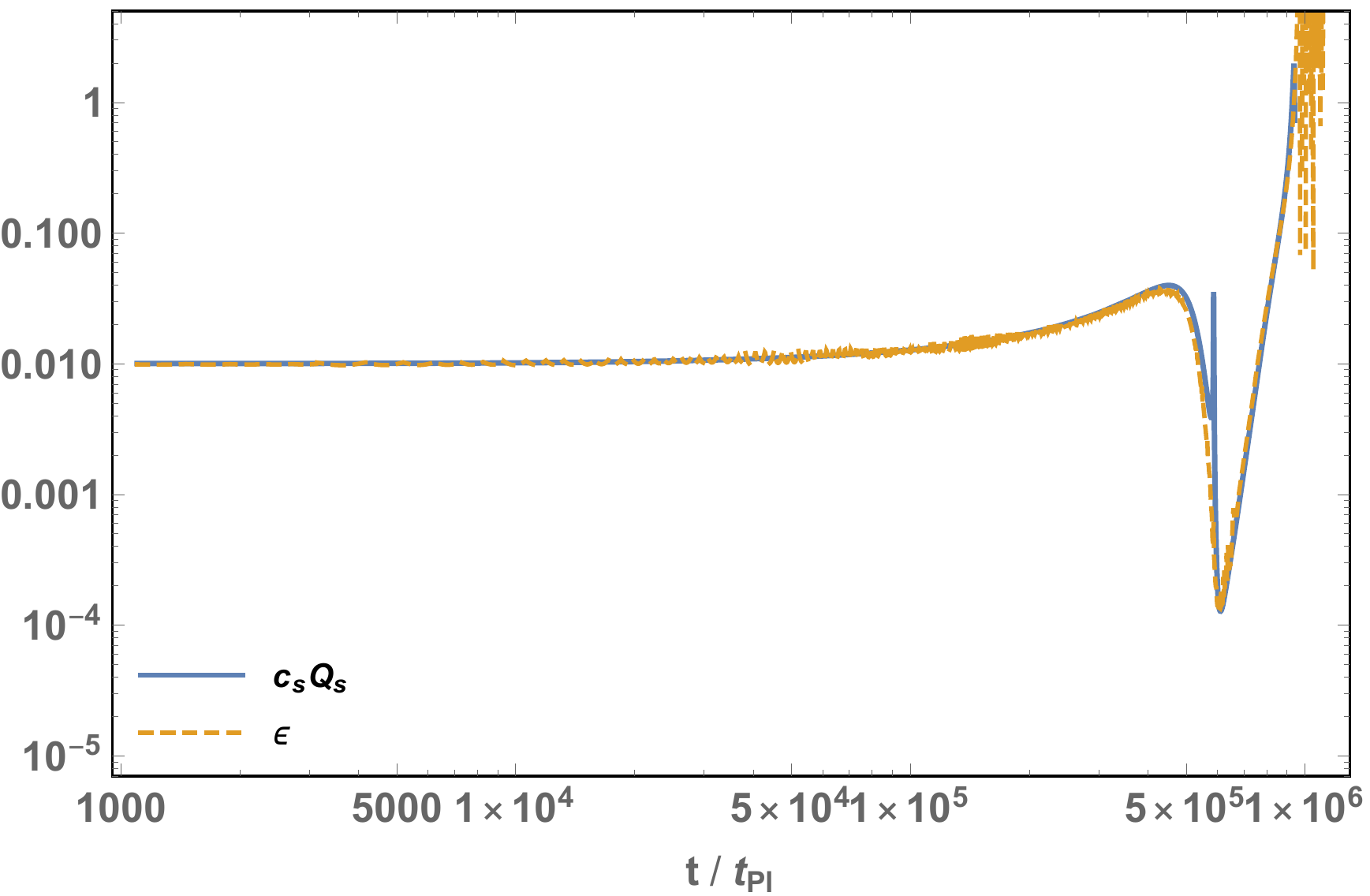}%
\hfill
\includegraphics[width=0.44\textwidth]{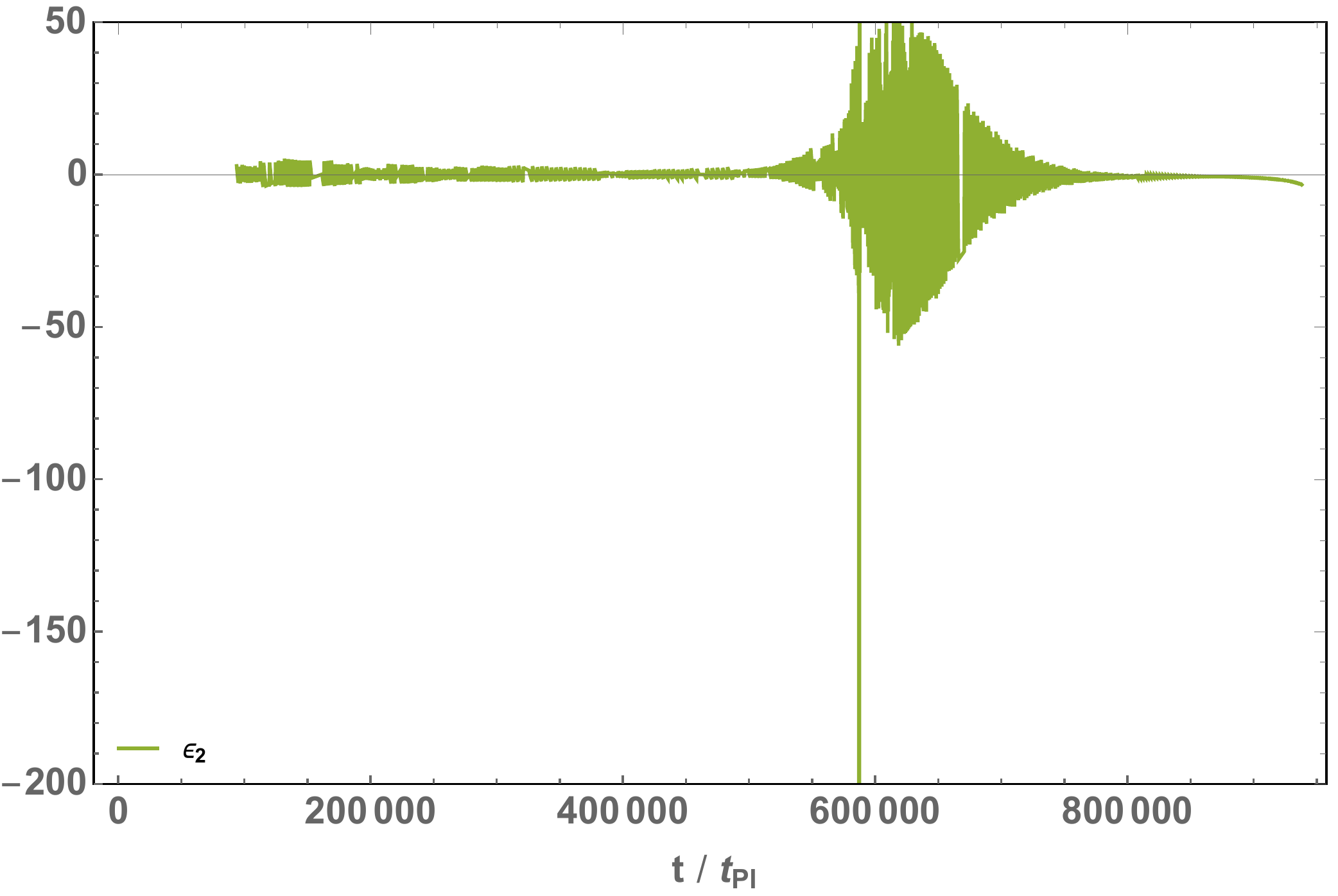}%
}\\[0.4cm]
\caption{{\it Left panel:} 
The $c_s Q_s$ (blue line) and the first Hubble flow parameter $\epsilon$ (orange dashed) are depicted in situations that the inflaton velocity decreases significantly. 
{\it Right panel:} The corresponding second Hubble flow parameter $\epsilon_2$ is depicted that becomes negative. The plots correspond to the case of $10^{21}$ grams PBHs.
 \label{Fig-Qcs}}
 \end{figure}

In conventional cases, the estimation of the power spectrum is done after omitting one of the two solutions of the Eq. (\ref{MS}).  However, in situations that the slow-roll parameters change dramatically in the course of the inflaton's evolution then this is not the case. 
This is easily seen after rewriting the Mukhanov-Sasaki equation in the form  
\begin{equation} \label{RMS}
{\cal R}_k''+(2+{\epsilon}_2)aH{\cal R}'_k+c^2_s k^2 {\cal R}_k=0.
\end{equation}
The $\epsilon_2 $ is the second Hubble-flow parameter, $\epsilon_2 \equiv \dot{\epsilon}/(H\epsilon)$. Eq. \eqref{MS} is recast into the form (\ref{RMS}) after the observation that $c_s Q_s \simeq \epsilon$ where $\epsilon=-\dot{H}/H^2$ is the first Hubble-flow parameter. This is apparent for $\epsilon_D\ll 1$, see Eq. (\ref{eD}), and $c_s\simeq 1$. It is also true for the  interesting cases that the velocity of the inflaton field decreases significantly and the power spectrum increases, as the figure \ref{Fig-Qcs} demonstrates. 

 At the large scale limit the last term is negligible and for $(2+{\epsilon}_2)>0$ one finds a constant and decaying mode for the curvature perturbation.  However if the parenthesis is negative the second solution corresponds to a growing mode and the omitted solution contributes significantly to the power spectrum. 
In our models, the first slow-roll parameter depends on the GNMDC function $f(\phi)$ and if $f(\phi)$ changes  abruptly the ${\cal R}$ can be enhanced with interesting implications for the PBH scenarios.

\section{PBH production from Higgs inflation with GNMDC}

Let us consider the inflationary potential $V(\phi)=(\lambda/4) \phi^4$, that can be identified with the Higgs field for $\lambda \simeq 0.1$.
As we have shown, the Higgs field can give a viable inflation if a GNMDC operates. Let us assume the non-minimal coupling to be of the form, (\ref{monom}) i.e $f_I(\phi)= {\alpha  \phi^{\alpha-1}}/{M^{\alpha+1}}$.
It is helpful to label ${\epsilon}_\text{PBH}\equiv \epsilon_V(\phi_0)/{\cal A}(\phi_0)$ the minimum value of the ratio $\epsilon_V/{\cal A}$ which is the quantity that modulates  the power spectrum amplitude; note also that $\epsilon(\phi_0) \simeq {\epsilon}_\text{PBH}$.
 For $f_{II}(\phi)$ given by eq. (\ref{fIIr}) 
the size of the parameter $d$, the  GNMDC "strength",  is given by the relation
\begin{align} \label{epbh}
d \sim \frac{M^2_\text{Pl}}{V(\phi_0) f_I(\phi_0)} \frac{\epsilon_V}{{\epsilon}_\text{PBH}} =\frac{8}{\alpha \lambda} \frac{M^4_\text{Pl} M^{\alpha+1}}{\phi_0^{\alpha+5}} \frac{1}{{\epsilon}_\text{PBH}}\,.
\end{align}
The number of $e$-folds $N_{II}\leq N_\text{PBH}$  take place during the second stage of inflation with $f(\phi) \simeq f_I(\phi) f_{II}(\phi)$. 
 For a particular $ N_{II}$ value the $s$ parameter of (\ref{fIIr}) can be specified.

 We comment that the exact values of the parameters are determined after solving the Mukhanov-Sasaki equation and by computing the power spectrum.
 The above relations are useful mostly for a qualitative approach nevertheless they make clear the underlying assumptions and physics. 
Below we present an explicit example. 
 
\begin{figure}[htbp]
\makebox[\textwidth]{
\includegraphics[width=0.47\textwidth]{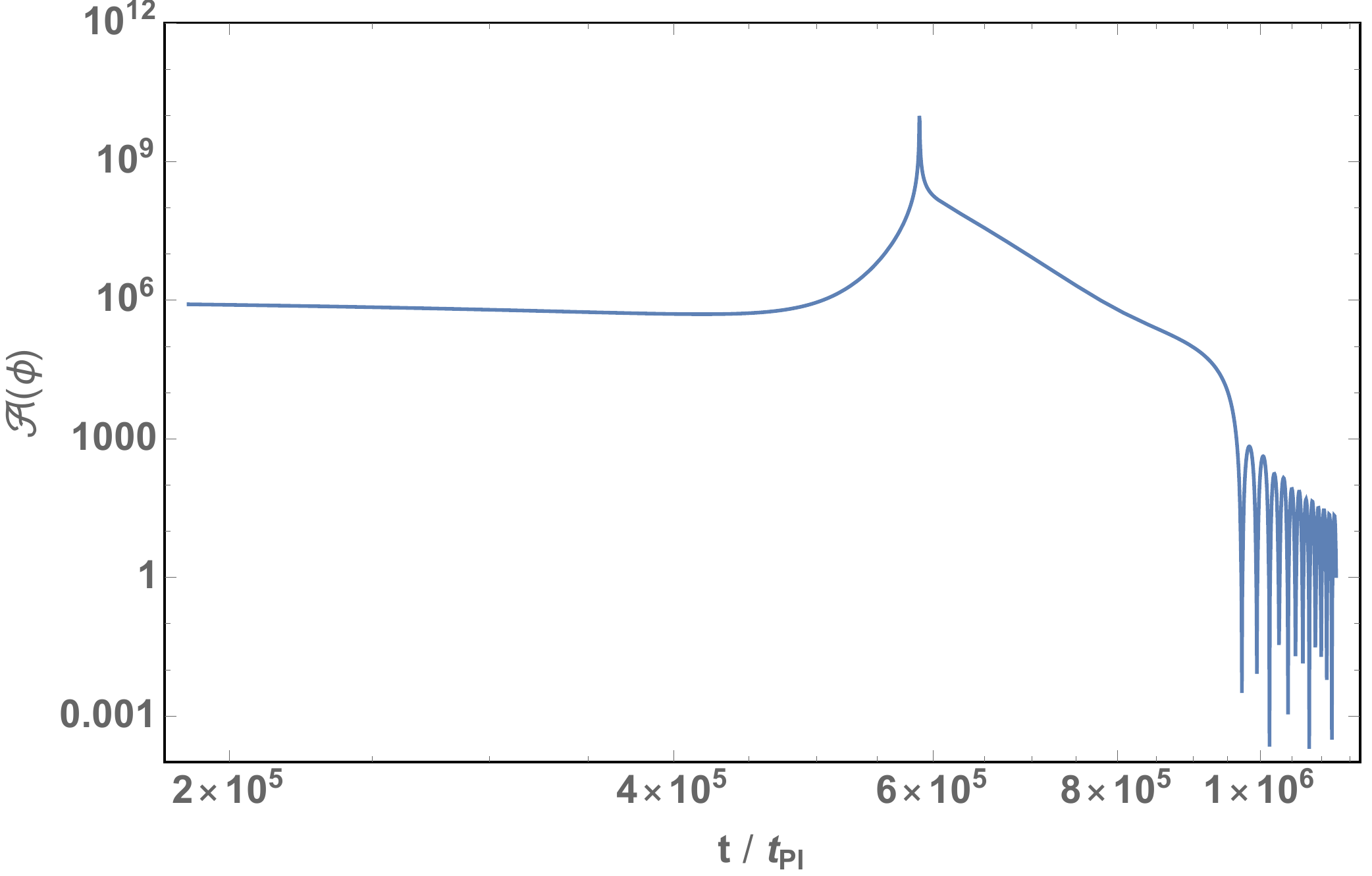}%
\hfill
\includegraphics[width=0.47\textwidth]{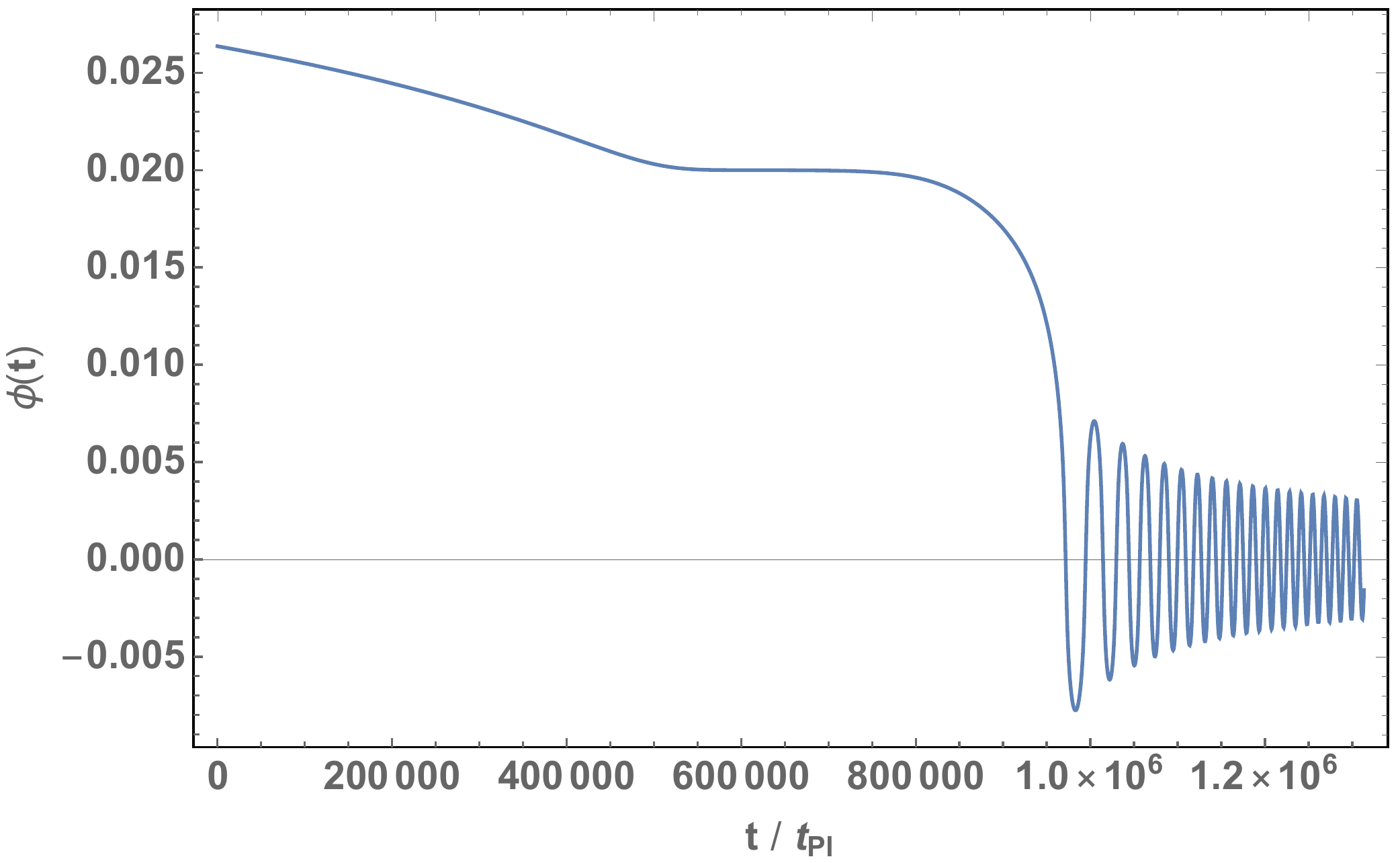}%
}\\[0.4cm]
\caption{{\it Left panel:} The ${\cal A}=1+3H^2 f(\phi)$ that realizes inflation with GNMDC and generates a peak sufficient to trigger PBH production with mass about $10^{21}$g. 
 {\it Right panel:} The evolution of the inflaton field value with GNMDC and standard model Higgs potential, $V(\phi)=\lambda \phi^4/4$.  The plateau is due to the enhanced friction that the $f_{II}(\phi)$ factor of the GNMDC  generates.
 The decrease of the inflaton velocity enhances the power spectrum at $k\sim  10^{12}$ Mpc$^{-1}$. }
  \label{A-21}
\end{figure}

\subsubsection{$10^{21}$ grams PBHs as dark matter}

According to Eq. (\ref{N2pbh}) a PBH with mass $M_\text{PBH }\sim 10^{21}$ g is generated if the amplitude of the scalar perturbations gets enhanced about $N_\text{PBH} \sim 27$ $e$-folds before the end of inflation. This number
gets modified if the reheating temperature is low \cite{Dalianis:2018frf, Dalianis:2018ymb}.
In our case it is a quartic potential, $\phi^4$ or Higgs inflation models, that drives the inflaton oscillations
hence the background expansion rate is equivalent to a radiation dominated universe with very large reheating temperature.
In addition the PBH abundance is cosmologically significant, if a mass fraction $\beta(M)\sim 10^{-13}$ of the universe  collapses into PBHs with $M_\text{PBH }\sim 10^{21}$ g mass.
The approximate relation  (\ref{bP}) tells us that the required amplitude of the power spectrum for such a scenario to be realized is  ${\cal P_R}\sim  10^{-2}$;
 and from Eq. (\ref{PR2}) the value of the ratio $\epsilon_V/{\cal A}\simeq {\epsilon}_\text{PBH}$ is approximated.

The inflationary dynamics and the parameter values are fully determined after fitting the CMB inflationary observables.
Higgs-like inflation, followed by a continuous thermal phase after reheating, predicts that the CMB scale $k_\text{CMB}=0.05$ Mpc$^{-1}$ exits the Hubble horizon about 58 $e$-folds before the end of inflation.
The total observable $e$-folds are partitioned between the two inflationary stages with $N_I$ and $N_{II}$ $e$-folds respectively.
Initially a number $\lesssim N_I$ of $e$-folds of inflation with $f_{I}(\phi)$ GNMDC take place, afterwards the friction increases due to $f_{I}(\phi)f_{II}(\phi)$ GNMDC that nearly immobilizes the inflaton for $N_{II}$ $e$-folds, and finally the  $f_{I}(\phi)$ might take over again for the remaining $N_I$ $e$-folds. The scalar tilt value determines the $N_I$.
The required $N_I$ value can be significantly reduced  when the running of the running is included  \cite{Akrami:2018odb}.
 In addition, for a particular $N_I$ the CMB normalization for Higgs infation with $\lambda\sim 0.1$ fixes the scale $M$ of the GNMDC.
 
In order to illustrate the GNMDC dynamics  we present an explicit example after solving the Mukhanov-Sasaki equation numerically. About 500 modes ${\cal R}_k$, are presented with a red dot in Fig. \ref{fPBH}. We find a power spectrum that features a peak high enough to generate a significant abundance of PBHs. The parameters of our example have values $\alpha=3$, $M=7.1 \times 10^{-5} M_\text{Pl}$, $d=5.5 \times 10^6$, $s=2.1\times10^{-10}$. They yield $N=50$ $e$-folds from the inflaton field value $\phi_\text{CMB}=0.0264 M_\text{Pl}$ until the value $\phi_\text{end}=0.0197 M_\text{Pl}$. At $\phi= 0.02 M_\text{Pl}$, the $f_I(\phi)f_{II}(\phi)$ term dominates and the inflaton velocity decreases dramatically. This is demonstrated by the fall of the $\epsilon$ parameter,  depicted in  Fig. \ref{Fig-Qcs}. At that point the power spectrum increases sharply reaching the value ${\cal P_R}(k_\text{peak})\sim2\times 10^{-2}$, see Fig. \ref{A-21} and \ref{fPBH}.
 The $f_{II}(\phi)$ term gradually vanishes, the infaton slow-rolls few more $e$-folds to the end of inflation. Spectral index values  in accordance with the Planck data can be found.  For these parameters, the PBH have mass roughly $10^{21}$ grams  and the total fractional abundance of the PBHs is found to be $f_\text{PBH} \sim 0.1$, see Fig. \ref{fPBH}. For the estimation of the PBH abundance we followed the Press-Schechter formalism presented in Section \ref{PrePBH} with threshold parameter $\delta_c \simeq 0.4$, a value close to the one suggested by Ref. \cite{Harada:2013epa}. 
 
 We comment that the numerical solution of the Mukhanov-Sasaki equation yields a  power spectrum that deviates significantly from the approximate analytic curve even at small wavenumbers $k$, where the first slow parameter $\epsilon$ does not change significantly, as the Fig. \ref{fPBH} demonstrates. This peculiar feature is justified by observing that the $\epsilon$ numerical value oscillates at that $k$-interval, see Fig \ref{Fig-Qcs}. Though the oscillations have tiny amplitude the frequency is large and the second Hubble-flow parameter becomes negative $\epsilon_2<-2$ before the major fall of the $\epsilon$ value.  Hence the power spectrum increases gradually from the small wavenumbers. Another remark is that the  power spectrum points after the peak appear scattered. We attribute this feature to numerical subtleties. Finally, we note that particularly  large values for the power spectrum ${\cal P_R}(k)$ can be achieved as the numerical practice indicates; the difficulty of finding viable models mainly lies at the fitting of the model parameters with the CMB observables at the large scales. We further comment on this issue in the next paragraphs.

\subsubsection{LIGO mass PBHs}

According to eq. (\ref{N2pbh}) a PBH with mass $M_\text{PBH }\sim 10^{35}$ g is generated if the amplitude of the scalar perturbations gets enhanced about $N_\text{PBH} \sim 45$ $e$-folds before the end of inflation.
In addition, the PBH abundance is cosmologically significant if a mass fraction $\beta(M)\sim 10^{-7}$ of the universe  collapses into PBHs with $M_\text{PBH }\sim 10^{35}$ g mass.
The approximate relation  (\ref{bP}) tells us that the required amplitude of the power spectrum for such a scenario to be realized is ${\cal P_R}\sim 1.1 \times 10^{-2}$; and from eq. (\ref{PR2}), the ratio $\epsilon_V/{\cal A}$ is approximately found.

Again, inflationary dynamics and the parameters' values are fully determined after fitting the CMB inflationary observables.
As before, Higgs-like inflation predicts that the CMB scale $k_\text{CMB}=0.05$ Mpc$^{-1}$ exits the Hubble horizon about 58 $e$-folds before the end of inflation. For such a large PBH mass, it is hard to find a working example with spectral index $n_s$ in accordance with the last Planck CMB data. Nevertheless, our practice with the numerics involved in the computation of the power spectrum does not rule out the possibility that a different set of parameters (or a different $f_{II}(\phi)$ function) might exist that yield a $n_s$ value compatible with the observational constraints.

\begin{figure}[htbp]
\makebox[\textwidth]{
\includegraphics[width=0.47\textwidth]{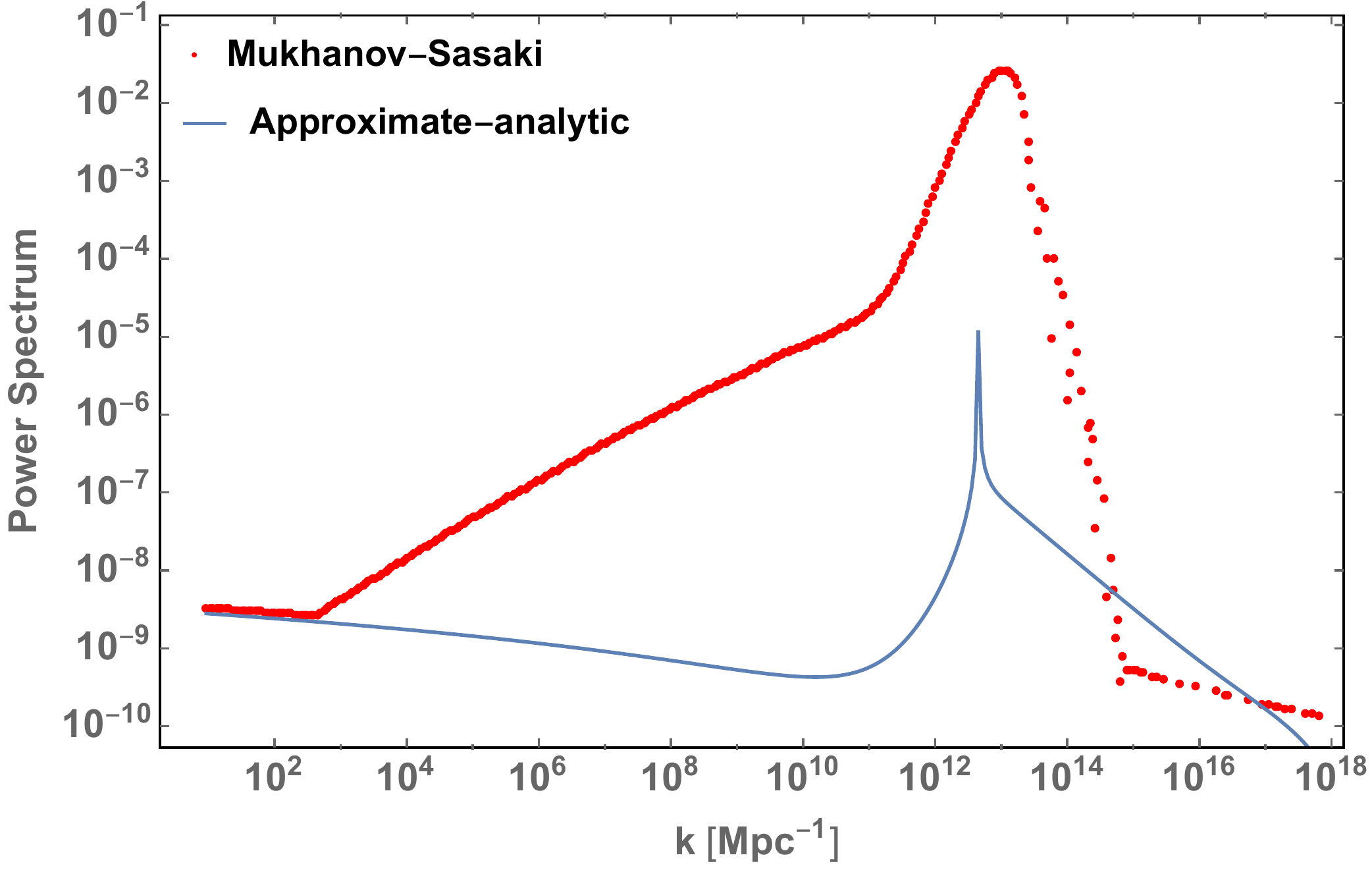}%
\hfill
\includegraphics[width=0.47\textwidth]{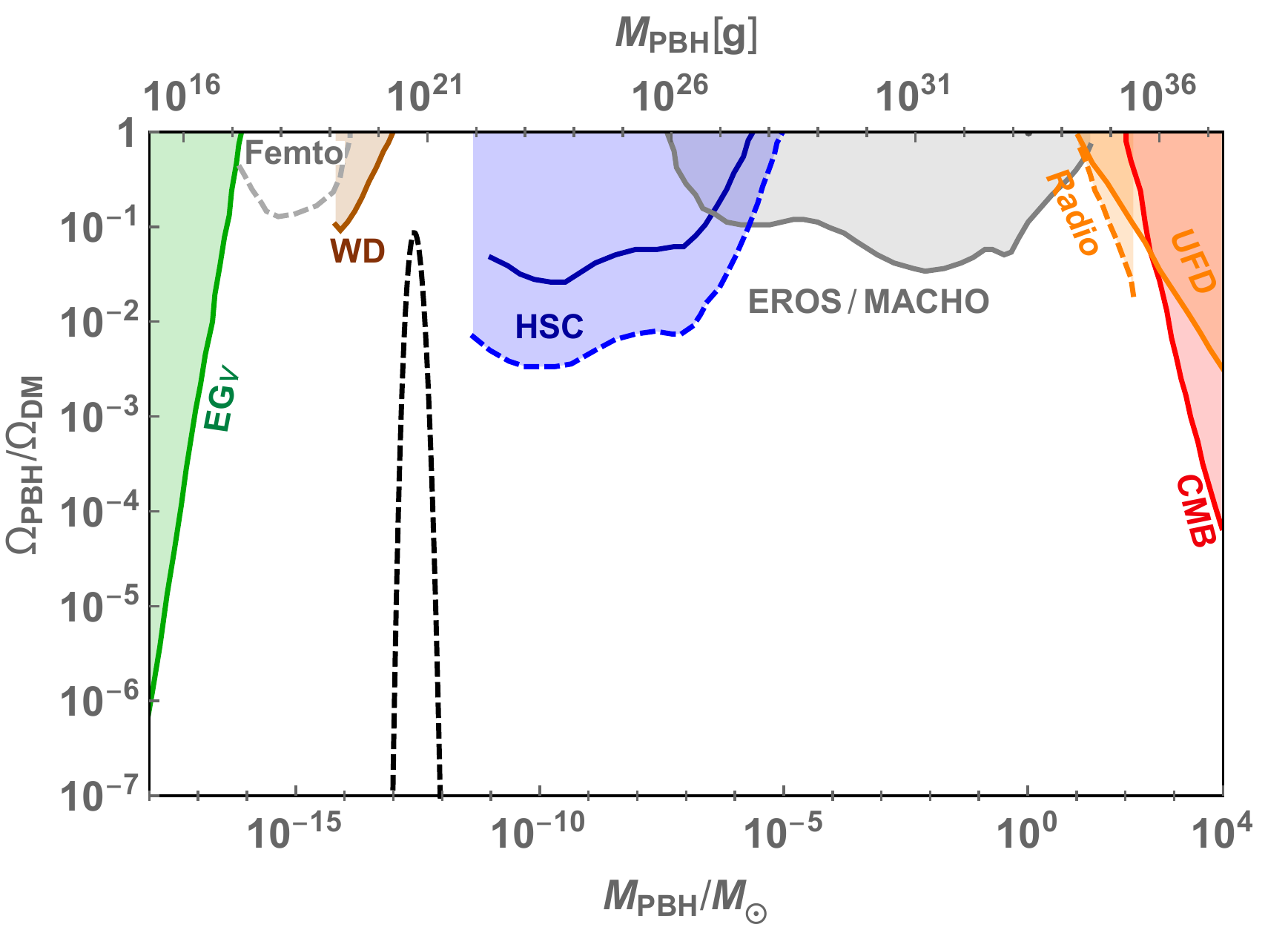}%
}\\[0.4cm]
\caption{
{\it Left panel}: The power spectrum of the curvature perturbation ${\cal P_R}(k)$ with parameters as described in the text,  due to  GNMDC and Standard Model  Higgs-like potential. The dotted red line depicts the power spectrum computed numerically from the exact Mukhanov-Sasaki equation while the solid line depicts the approximate expression (\ref{PS-SR}).
{\it Right panel}: The corresponding fractional abundance of the PBHs,  $f_\text{PBH}\sim 0.1$,  with central mass $M_\text{PBH}\sim 10^{21}$ g together with the observational constraints is depicted.
}
  \label{fPBH}
\end{figure}

\subsubsection{PBHs remnants as dark matter}

Let us discuss briefly here the scenario with mini and promptly evaporating PBHs  in the context of the GNMDC, since PBHs of certain masses could be a viable dark matter candidate \cite{Belotsky:2014khl}.
It can be realized if the 
 the power spectrum features a peak at the smallest scales or it becomes blue towards the end of inflation.
In this case the generated PBHs  are ultra light, and are expected to evaporate promptly without affecting the big bang observables, such as the BBN and the CMB.
There are several theoretical reasons to anticipate that PBHs do not evaporate to nothing, but instead leave behind a stable mass state, called PBH remnant, with mass $M_\text{rem}=\kappa M_\text{Pl}$,  see e.g \cite{Barrow, Carr:1994ar}. Here, $\kappa$ is a factor that parameterizes our ignorance  about the physics that operates at the Planck scales.
A very attractive feature of this scenario is that the spectral index value $n_s$ can be  inside the $68 \%$  CL contour region of Planck 2018 data. 

In Ref. \cite{Dalianis:2019asr} it has been shown that power spectrum values ${\cal P_R} \sim 10^{-3} -10^{-2}$ can generate a population of PBH remnants with significant abundance, sufficient to explain all of the dark matter density in the universe for a wide range of $\kappa$ values. The fractional abundance of the PBH remnants with respect to the total dark matter has been found for general background expansion rates. In the case that PBHs are produced during the radiation domination stage it is
\begin{equation} \label{fRD2}
f_\text{rem} (M)\, \simeq\,\kappa \left(\frac{\beta_\text{}}{10^{-12}}\right) \,
 \Big(\frac{\gamma_\text{}}{0.2}\Big)^{\frac{3}{2}}
\left(\frac{M}{10^{5}\text{g}}\right)^{-3/2}\,,
\end{equation}
where $M$ is the mass of the parent PBH.
Large field inflation models can admit PBH values as small as $M\sim 10$ g and hence $\beta$ values as small as  $10^{-18}$ are possible to generate a significant PBH abundance.

Actually, it is numerically easier to construct a viable PBH inflationary model if the power spectrum peak is at the very end of inflation. The reason is that a ${\cal P_R}(k)$ with a peak at the smallest scales is less constrained by the data (microlensing, Hawking radiation, etc) and, also, acceptable $n_s$ values are easier obtained. 
Additionally, a power spectrum that turns from red into blue  might also generate mini PBHs. 
 A GNMDC that increases steadily in the course of inflation can realize this scenario. 
An example considered in this paper,  is the coupling function $f(\phi) =M^{-2} \, e^{-2\tau \phi}$, which for particular $\tau$ values, can trigger the generation of mini PBHs.
The inflationary potential has to be accordingly chosen, so that the spectral index value is in agreement with last data as described in Section \ref{models}.
An interesting implication of the PBH remnant scenario is that inflationary potentials without a minimum are phenomenologically acceptable.
The reason is that the reheating of the universe can be realized via the PBH evaporation; the cosmology of this scenario  was described in Ref.  \cite{Dalianis:2019asr}.

We note that the increase of the GNMDC strength towards the end of inflation might be problematic during the reheating stage; see the discussion in section \ref{models},
where the idea of a vanishing $f(\phi)$  after the end of inflation was put forward.  
An alternative way to address the reheating instabilities is to introduce an extra dimension in the field space in such a way that  the system does not oscillate around the $\phi$ direction.
These hybrid inflation models \cite{Linde:1993cn},
\begin{align} \label{hybr}
V_\text{hyb}(\phi, \chi)=V_0(1-\chi^2/\mu^2)^2+V(\phi) +\frac{g^2}{2}\chi^2 {\phi^2}~,
\end{align}
provide a solution.   
They can be considered effectively as single field models for the inflaton $\phi$, where the $V(\phi)$ can be the Higgs potential. 
 Inflation ends either due to the end of the slow-roll, or by the waterfall transition of $\chi$ at $\phi=\phi_c$.  The former case is realized when
 $V(\phi) \gtrsim V_0$
  whereas the latter for $V(\phi) \ll V_0$.
After inflation it is the minimally coupled $\chi$  field that oscillates and not the inflaton $\phi$ which gets stabilized at $\phi=0$.
Assuming that the GNMDC increases towards the end of inflation, then, as $\phi \rightarrow \phi_c$, the inflaton field decelerates and the power spectrum gets  enhanced.  Right before the  $\phi_c$ we expect the minimum velocity at the $\phi$ direction and thus the maximum value for the ${\cal P_R}(k)$.
  For $\phi>\phi_c$ a waterfall transition takes place, the system  transits towards the $\chi$ direction and a $\chi$ field oscillating phase takes over inflation.

  Summarizing, enhancement of the power spectrum at ultra small scales $k^{-1}$,  that realize  the PBH remnant dark matter scenario, is well possible inside the framework of the GNMDC. Possible  reheating instabilities, due to residual GNMDC,  can be circumvented   if extra field directions exist.

\section{Conclusions}
\label{SecConc}

The inflationary paradigm is widely accepted to be a viable phenomenological framework providing the initial conditions for the hot big bang and describing the generation of primordial perturbations.  Inflation is realized either by non-steep  scalar field potentials or by particular  non-canonical kinetic terms.
In this work we investigated the cosmology of the ${\cal L}_5$ Horndeski term, that introduces a non-minimal derivative coupling (NMDC) of the inflaton field to the Einstein tensor. The cosmology of this coupling has been studied in several works in the past, but most of them studied the simplest, field independent version of the coupling, whereas in the present work a more general, field dependent NMDC was considered, that we dubbed GNMDC.

Similarly to other classes of inflationary models, such as the DBI
or $\alpha$-attractor models,
GNMDC inflation features  a non-canonical kinetic term that acts like friction, decelerating the inflaton field and implementing a slow-roll phase,  even when the potential is steep. The field dependence of the GNMDC produces a new and richer inflationary phenomenology. In addition, inflation with GNMDC can be free from possible gradient instabilities during the postinflationary oscillating phase, if the coupling is chosen to vanish at the bottom of the inflationary potential.

In this work we investigated the inflationary phenomenology of the GNMDC, deriving the relations for the spectral index and the tensor-to-scalar ratio.
It was known that the sound speed squared of the scalar perturbation rapidly oscillates between positive and negative values and an instability for the shortest wavelength mode of the scalar perturbation may take place making  the dynamics of the system difficult to follow  analytically.
  We put forward viable forms for the GNMDC in order to ameliorate (see eg. Fig. \ref{fig:sspdcomparison}) or completely avoid instabilities during the reheating stage (see eg. Eq. (\ref{hybr})) and described inflation with simple and motivated potentials.
In particular, in the framework of the GNMDC we examined the so called new-Higgs inflation \cite{Germani:2010gm}, which is rather exciting since the Higgs field is the only scalar discovered in nature \cite{Chatrchyan:2012xdj}.  We also examined the exponential potential that can successfully  drive inflation when the GNMDC operates, in agreement with the Planck 2018 data \cite{Akrami:2018odb}.
Moreover, inflation with exponential potentials and GNMDC ends naturally, due to the gradual decay of the GNMDC term and a kination regime might follow.

A correspondence between the inflaton dynamics with GNMDC and the dynamics of a canonical inflaton with GR gravity has been described. This correspondence makes the prediction of the complicated GNMDC dynamics easily understood and anticipated. It also makes model selection and identification of GNMDC dynamics through the CMB observables possible.

The fact that GNMDC is modulated by the field value $\phi$ implies that the friction term can be enhanced during the course of the inflationary stage. The enhancement of the friction slows down  the inflaton field  and amplifies the amplitude of the curvature perturbations.
If the power spectrum is sufficiently amplified then PBHs can be produced thanks to the GNMDC, within single field inflation models.
The power spectrum amplification is achieved by choosing a tailor-made coupling of the inflaton to the Einstein tensor. Nevertheless, the inflaton potential can be very simple.
In this work we implemented the PBH production using the Standard Model Higgs potential.
Being illustrative, we estimated the PBH abundance for benchmark cases using a semi-analytic methodology. We focused on the particularly interesting PBH mass window of $M\sim 10^{21}$g  that can explain the dark matter density in the universe and we presented numerical results.
We also commented on  the scenario that the  GNMDC  increases steadily towards the end of inflation triggering the production of mini PBHs, hence PBH remnants might be produced.
 Our results are suggestive underlining the possibilitites that the GNMDC  introduces to  the inflationary cosmology. We did not discuss possible relevant issues such as quantum diffusion, or non-Gaussian effects.  

Summarizing, in this work we put forward inflation models with GNMDC that  give new and distinguishable inflationary predictions in a more reliable framework, where the GNMDC vanishes fast after inflation. Also, we further elaborated the implications of the inflaton-modulated GNMDC and constructed models that can adequately amplify the power spectrum of primordial perturbations at small scales, triggering PBH production. An attractive feature of  the GNMDC is that inflation as well as PBH production can be implemented utilizing the Higgs or a Higgs-like potential.


\section*{Acknowledgments}
We thank S. Tsujikawa for discussions during the early stages of this work.
We would like also to thank C. Germani and P. Wu  for correspondence.
The work of I.D. is supported by IKY Scholarship, co-financed by Greece and the European Union (European Social Fund-ESF), through the Operational Program "Human Resources Development, Education and Lifelong Learning" in the context of the project “Reinforcement of Postdoctoral Researchers - 2nd Cycle” (MIS-5033021), implemented by the State Scholarships Foundation.

\end{document}